\documentclass[11pt]{article}

\usepackage{hyperref}
\usepackage{amssymb}
\usepackage{amsmath}
\usepackage{epstopdf}
\usepackage{morefloats}
\usepackage{graphicx}
\usepackage{hyperref}
\usepackage{subfigure}
\usepackage{subfigmat}
\usepackage{bm}
\usepackage[all]{nowidow}
\usepackage{color,soul}
\usepackage[dvipsnames]{xcolor}
\usepackage{caption}
\usepackage{verbatim}
\usepackage{mwe}
\newcommand{\inner}[2]{\big< #1 ,  #2 \big>}
\usepackage[hmargin=3cm]{geometry}
\usepackage{placeins}
\usepackage[numbers,square,sort&compress]{natbib}
\usepackage[noabbrev]{cleveref}
\usepackage{todonotes}
\newcommand{\bs}[1]{\ensuremath{\boldsymbol{#1}}}

\usepackage[affil-it]{authblk}
\usepackage[font=scriptsize]{caption}

\begin{document}

\title{On-the-fly Reduced Order Modeling of  Passive and Reactive Species via Time-Dependent Manifolds}
\author[1]{Donya Ramezanian}
\author[1]{Arash G. Nouri}
\author[1]{Hessam Babaee\thanks{Corresponding author. Email:h.babaee@pitt.edu.}}
\affil[1]{\footnotesize Department of Mechanical Engineering and Materials Science, University of Pittsburgh}
% \affil[2]{\footnotesize NASA Langley Research Center, Hampton, VA}
% \and Mark Carpenter\thanks{NASA Langley Research Center, Hampton, VA
%   (\email{mark.h.carpenter@nasa.gov}).}
% \and Hessam Babaee\footnotemark[2]}
\date{}
\maketitle

% \title{On-the-fly Reduced Order Modeling of  Passive and Reactive Species via Time-Dependent Manifolds}

%  \author{
%   Donya Ramezanian,  Arash G. Nouri, Hessam Babaee\thanks{Assistant Professor; email: h.babaee@pitt.edu} \\
% {\normalsize\itshape
%          Department of Mechanical Engineering and Materials Science,
%          University of Pittsburgh, 3700 O’Hara Street, Pittsburgh, PA 15213, USA, \\
%  }
%  }
 
% \cortext[mycorrespondingauthor]{Corresponding author. \\
% E-mail addresses: h.babaee@pitt.edu}

%%%%%%%%%%%%%%%%%%%%%%%%%%%%%%%%%%%%%%%%%%%%%%%%%%%%%%%%%%%
\begin{abstract}
 One of the principal barriers in developing accurate and tractable predictive models in turbulent  flows with a large number of species is  to track every species by solving a separate transport equation, which can be computationally impracticable.   In this paper, we present an \emph{on-the-fly} reduced order modeling  of   reactive as well as passive transport equations  to reduce the computational cost.  The presented approach seeks a low-rank decomposition of the species to three time-dependent components: (i) a set of  orthonormal spatial modes, (ii) a low-rank factorization of the instantaneous species correlation matrix, and (iii) a set  of orthonormal species modes, which  represent a low-dimensional \emph{time-dependent manifold}.  Our approach  bypasses the need to  solve the full-dimensional species to generate high-fidelity data --- as it is commonly performed in data-driven dimension reduction techniques such as the principle component analysis. Instead, the  low-rank components are directly extracted from the species transport equation. The evolution equations for the three components are obtained from optimality conditions of a variational principle.  The time-dependence of the three components  enables an on-the-fly adaptation of the low-rank decomposition     to transient changes in the  species. Several demonstration cases  of reduced order modeling of passive and reactive transport equations are presented. 
 
\end{abstract}

\section{Introduction}
Tractable numerical simulation of passive and reactive species transport in turbulent flows has been the subject of intense research in the past few decades \cite{LL09}  due to its importance in a diverse range of engineering and scientific applications. See for example \cite{johnson1995,warhaft2000,anand2003,antonia2003,SM06,li2009,wang2015,cerrolaza2017}. 
% Examples include  heat transfer in air jet \cite{warhaft2000}, dispersion of pollutants in the atmosphere \cite{johnson1995,li2009},  plasma physics \cite{antonia2003}, and chemical mixing in  industrial \cite{wang2015} and biological processes \cite{cerrolaza2017, anand2003}. %Computational cost of direct numerical simulation (DNS) of all length and time scales associated with these problems is extravagantly high.
% Therefore, numerous studies have been conducted on how to model passive species fields  as well as ADR processes in chaotic and turbulent flows \cite{babiano1987,kenjerevs2015}. Development of validated, predictive, multi-scale mixing models for these flows require realistic simulations with realistic compositions.
High fidelity numerical simulations  using direct numerical/large-eddy simulation (DNS/LES)  of turbulent reactive flows with a large number of  species is cost prohibitive~\cite{P13}. Detailed kinetic models usually contain hundreds of species and thousands of reactions~\cite{SYWL09,L07,coltrin2003,NGL19}. Each species requires solving a transport equation, i.e., a partial differential equation (PDE). The cost of solving a large number of transport PDEs on a DNS/LES grid can be prohibitive~\cite{CFD2030}. Moreover, the memory and input/output (I/O) cost of storing a large number of species imposes severe limitation on the number of species that can be simulated.  For example, currently I/O limitations allow for the storage of the species at every 400 simulation time steps in a DNS of turbulent combustion, while intermittent phenomena such as ignition kernel occurs in the order of 10 simulation time steps \cite{BAB12}. The cost of I/O and storage continues to grow in comparison to the cost of floating point operations in the future  high performance computing architectures and this trend will impose even more stringent I/O  and memory restrictions  \cite{AM_DOE_14}.  

To reduce the computational cost of large scale complex turbulent reactive simulations,   a vast array of  techniques have been developed with the aim of reducing the number of species and/or reactions. Skeletal reduction, for example~\cite{LL09}, finds an optimized subset of a detailed model by eliminating unimportant species and reactions. The skeletal reduction is often performed for zero-dimensional reactors and based on different criteria, e.g. sensitivity analysis~\cite{VT86,EC11,SFCFR16}, reaction flux analysis~\cite{lu2005,pepiot2008,niemeyer2010}, etc.~\cite{Elliott04,Sikalo14}. Other reduction techniques include lumping methods~\cite{Djouad03,Gao16} and time-scale analysis techniques e.g. quasi steady state approximation~\cite{Stiefenhofer98}, partial equilibrium approximation~\cite{Rein92}, rate controlled constrained equilibrium~\cite{Keck90}, computational singular perturbation~\cite{LG89,GIV13} and intrinsic low dimensional manifold~\cite{MP92}. %In lumping, similar species and reactions are combined to reduce the number of variables to be tracked, while in time-scale analysis the concentration is on stiffness removal. 
% In general, all the reduction techniques are either based on local analysis of mixture state over a limited range of conditions (\textit{i.e.} species concentrations) in skeletal models or they track each species in time with a separate transport equation.

Reduction schemes based on principle component analysis (PCA)~\cite{SP09, ME15,OE17,MOCP20} is another strategy to enable realistic high fidelity simulations of turbulent flows with many species. The  PCA-based reductions belong to \emph{data-driven} reduction techniques, in which a reduced composition space is obtained by performing PCA  on a DNS/LES dataset and then solve transport equations for the reduced principal components.   In the PCA workflow, one  must at least solve  one full-dimensional DNS/LES  \textit{a priori} and store the data. This step is done \emph{offline} and one of its implications is that the PCA-based reductions are fined-tuned for a target problem and in general there it is difficult to guarantee that the reduced composition space is accurate under  if operating conditions such as  boundary conditions, geometry, Reynolds number and Mach number change.   This motivates for an \emph{on-the-fly} reduction scheme.

Recently, new \emph{model-driven} dimension reduction techniques  have been introduced. in which the low-rank structure is extracted from the model. Dynamically orthogonal (DO) \cite{SL09},  bi-orthogonal (BO)    \cite{CHZI13} and dynamically bi-orthonormal decompositions (DBO) \cite{PB20}   are  model-driven stochastic reduced-order modeling techniques. For linear deterministic systems,  optimally time-dependent (OTD) reduction was introduced \cite{Babaee_PRSA}.  In all of these  techniques, closed-form (partial) differential  equations for the evolution of the low-rank structures (modes) are derived. Leveraging the mathematical elegance of these techniques, for some cases,  numerical and dynamical system analyses are used to shed light into their performance. See, for example,  \cite{CSK14} for the equivalence of DO and BO;   \cite{PB20} for the equivalence of DBO with BO/DO; \cite{KL07,MNZ15} for a  theoretical error  bounds for the approximation error;  \cite{BFHS17}  for the exponential convergence  of $r$ OTD modes  to the $r$ most dominant Lyapunov vectors of a dynamical system; \cite{BFHS17}  for a variational principle for deriving data-driven DO evolution equations.

In this paper, we present a new DBO formulation for on-the-fly low-rank decomposition of species transport equation. In particular, we present a novel variational principle for the extraction of the components of the low-rank decomposition. The reduction is achieved by extracting instantaneous correlations between different species on the fly and directly from the species transport equation without the need to generate data. We demonstrate the performance of the presented method for three cases.    The remainder of this paper is organized as  follows: In \S\ref{Methodology}, we present a variational principle for the determination of the DBO components. In \S\ref{Demonstration Cases}, we illustrate the performance of the presented  method by solving passive species transport equations with one thousand species. We also apply the presented method  to solve incompressible and compressible  reactive  flows. Finally, a brief summary of the present work is presented in \S\ref{Summary}.

%%such as directed relation graph (DRG) or computational singular perturbation (CSP).
%%such as PCA, MARS or Isomap

\section{Methodology}\label{Methodology}
\subsection{Definitions and Notation}
Our target problem is the passive/reactive transport equation given by:
\begin{equation}\label{eq:ADR}
\frac{\partial \Phi}{\partial t} + (v \cdot \nabla )\Phi =  \nabla \cdot ( \nabla \Phi \bs{\alpha})+ S(\Phi,\rho,  T),
\end{equation}
and augmented with appropriate initial and boundary conditions. Here, $x \in D \subset \mathbb{R}^d$ is the spatial coordinate, where $d=1,2$ or 3 is the dimension of the physical domain and $t$ is time. In Eq.\ (\ref{eq:ADR}), $v$ is the velocity vector, $\Phi = [\phi_1(x,t), \phi_2(x,t), ..., \phi_{n_s}(x,t)]$ is the species concentration, and  $S(\Phi,\rho,T)$ is the  source term, where $\rho$ is the density and $T$ is the temperature. For clarity in the exposition, we  use the \emph{quasimatrix} notation as introduced in \cite{bt2004}, in which one of the matrix dimensions,  representing the continuous space, is shown with $\infty$ and the other dimension is defined with an integer. Following this notation $\Phi(x,t) \in \mathbb{R}^{\infty \times n_s}$ and $S(\Phi,\rho,T) \in \mathbb{R}^{\infty \times n_s}$. Moreover, $\bs{\alpha}=\mbox{diag}\{ 1/Re Sc_1, 1/Re Sc_2, \dots, 1/Re Sc_{n_s}\}$ is a diagonal matrix of size $n_s\times n_s$ at each physical point $x$, where, $Re$ is the Reynolds number, and $Sc_i$ is the Schmidt number of the $i$th species. We derive the equations for the generic case where $\bm{\alpha}=\bm{\alpha}(T(x,t))$ is space-time variable due to its dependence to temperature. The term $(v \cdot \nabla )\Phi$ is interpreted as a quasimatrix of size $\infty \times n_s$: $(v \cdot \nabla )\Phi=[(v \cdot \nabla )\phi_1, (v \cdot \nabla )\phi_2, \dots, (v \cdot \nabla )\phi_{n_s} ]$. In case of $S(\Phi,\rho,T)=0$, Eq.\ (\ref{eq:ADR}) represents the passive transport equation, and for incompressible flow the source term is only a function of species, \textit{i.e.} $S=S(\Phi)$. 

  %Also,    $\boldsymbol \alpha = \mbox{diag}[\alpha_1, \alpha_2, \dots \alpha_{n_s}]$   is the diagonal matrix of diffusivity.  In the above notation,  $v \cdot \nabla \Phi=[ v \cdot \nabla \phi_1, v \cdot \nabla \phi_2, \dots, v \cdot \nabla \phi_{n_s}]$ and $\nabla ^2\Phi \boldsymbol \alpha$ is a matrix-matrix multiplication. 
  
  We define an inner product in the spatial domain between two fields $u(x)$ and $v(x)$  as:
\begin{equation}\label{eq:Inner Product}
\inner{u(x)}{v(x)} = \int_{D} u(x)v(x)dx,
\end{equation}
 and the  $L^2$ norm induced by this inner product as:
\begin{equation}\label{eq:L2}
\big\|u(x)\big\|_2 = \inner{u(x)}{u(x)}^{\frac{1}{2}}.
\end{equation}
The Frobenius norm of a quasimatrix $A(x)=[a_1(x), a_2(x), \dots, a_n(x) ] \in \mathbb{R}^{\infty \times n}$ is defined as:
\begin{equation}\label{eq:L2}
    \big\|A(x)\big\|_{\mathcal{F}} ^2= \sum_{i=1}^n \int_{D} a_i^2(x) dx.
     \end{equation}
    We also  define the inner product between quasimatrices $U(x,t)\in\mathbb{R}^{\infty\times m}$ and $v(x,t)\in \mathbb{R}^{\infty\times n}$  as 
\begin{equation*}
    S(t) = \inner{U(x,t)}{V(x,t)},
\end{equation*}
where $S(t)\in\mathbb{R}^{m\times n}$ is a matrix with components $S_{ij}(t) = \inner{u_i(x,t)}{v_j(x,t)}$, where $v_j(x,t)$ is the $j$th column of $V(x,t)$. Following the above definition, we observe that $\inner{UR_U}{V}=R_U^T\inner{U}{V}$ and $\inner{U}{VR_V}=\inner{U}{V}R_V$ for any $R_U \in \mathbb{R}^{m \times m}$ and $R_V \in \mathbb{R}^{n \times n}$.  

% We also define:
% \begin{equation}\label{eq:Mean}
% \bar{\Phi}(x,t) = \sum_{i=1}^{n_s} {\phi_i}(x,t),
% \end{equation}
%  the \emph{two-point correlation operator} of species at time $t$:
% \begin{equation}\label{eq:Cov1}
% C_x(x,{x}',t) = \overline{\Phi(x,t)\Phi({x}',t)} = \sum_{i=1}^{n_s} \phi_i(x,t)\phi_i(x',t).
% \end{equation}
%  and similarly, \emph{species correlation matrix}:
% \begin{equation}\label{eq:Cov2}
% C_{\phi_{ij}}(t) = \inner{\phi_i(x,t)}{\phi_j(x,t)}, \quad i,j=1,\dots, n_s,
% \end{equation}
%  are defined with Eqs.\ (\ref{eq:Mean}-\ref{eq:Cov2}).
\subsection{Dynamically Bi-orthonormal Decomposition}
\begin{figure}[tb!]
\begin{center}
 \includegraphics[width=1\textwidth]{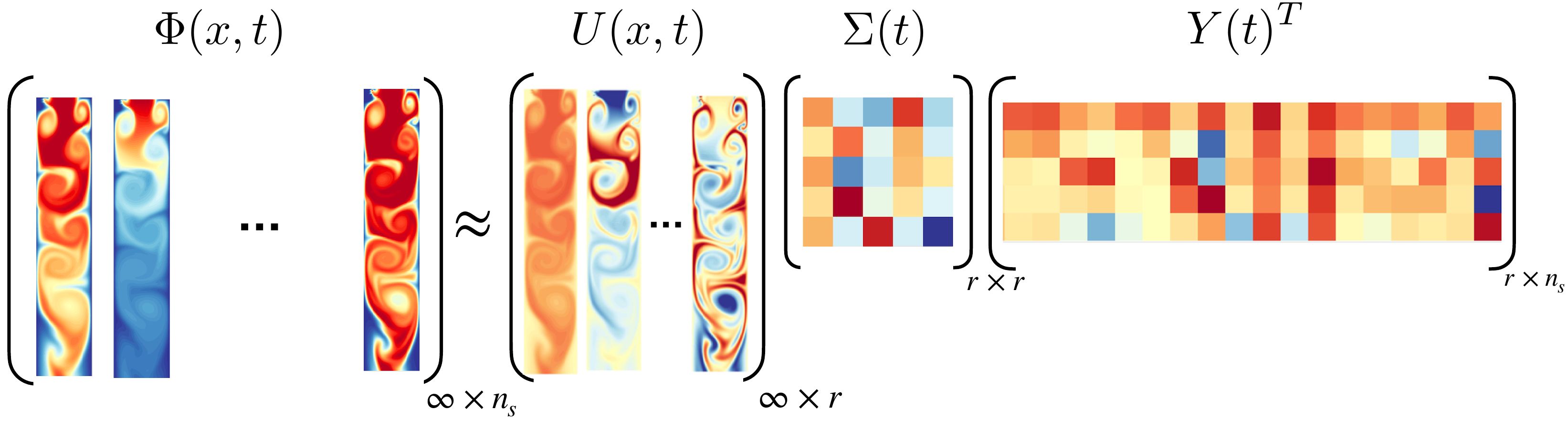}
 \caption{Schematic of dynamically bi-orthonormal  decomposition of species. The species quasimatrix $\Phi(x,t)$ is decomposed to  $U(x,t) \Sigma(t) Y(t)^T$, where $U(x,t)$ represents a set of orthonormal spatial modes, $\Sigma(t)$ is a factorization of the low-rank correlation matrix and $Y(t)$ is the matrix orthonormal species modes and it represents a low-dimensional time-dependent manifold. In the above decomposition, $n_s$ represents the number of species, $r$ is the rank of DBO and in the discrete representation $\infty$ will be replaced with the total number of grid points.        }
 \label{fig:DBO-Demo}
 \end{center}
\end{figure}

Our goal is to solve for a low-rank decomposition of  $\Phi(x,t)$ instead of solving Eq.\ (\ref{eq:ADR}).   To this end, we consider the DBO decomposition for the full species as  follows:
\begin{equation}\label{eq:DBO}
\Phi(x,t) = \sum_{j=1}^{r} \sum_{i=1}^{r} u_i(x,t)\Sigma_{ij}(t)y^T_j(t) + E(x,t),
\end{equation}
 where
\begin{subequations}\label{eq:USigmaY}
\begin{align}
U(x,t) &= [u_1(x,t)\mid u_2(x,t)\mid ...\mid u_r(x,t)],\\
Y(t) &= [y_1(t)\mid y_2(t)\mid ...\mid y_r(t)].
\end{align}
\end{subequations}
  Here, $r<n_s$ is the reduction size, $U(x,t)\in \mathbb{R}^{\infty \times r}$ is a quasimatrix whose columns are orthonormal spatial modes, $\Sigma(t)\in \mathbb{R}^{r\times r}$  is a reduced factorization  of the correlation matrix, $Y(t)\in \mathbb{R}^{n_s \times r}$ is the matrix of  orthonormal species  modes and $E(x,t) \in \mathbb{R}^{\infty \times n_s}$ is the reduction error (Fig.\ \ref{fig:DBO-Demo}). The key observation about the above decomposition, given by Eq.~(\ref{eq:DBO}) is that all three components are time dependent. As we explain, the time dependence of   DBO components  enables the decomposition to adapt on the fly to changes in $\Phi(x,t)$.  Orthonormality of spatial modes and  species modes implies that:
\begin{subequations}\label{eq:Orthonormality}
\begin{align}
\inner{u_i(x,t)}{u_j(x,t)} =\delta _{ij}, \label{eq:U-Ortho}\\ 
y^T_i(t)y_j(t) =\delta_{ij}  \label{eq:Y-Ortho}.
\end{align}
\end{subequations}

The schematic of the DBO decomposition is shown in Figure \ref{fig:DBO-Demo}. In the cases, where $\Phi(x,t)$ represents mass fraction, $n_s-1$ species can be used as the mass fraction of one of the species can be computed from $\sum_{i=1}^{n_s} \phi_i(x,t)=1$. Moreover, it is possible to include temperature and pressure in DBO in a straightforward manner. However, to maintain the generality of the presented approach for passive transport and incompressible reactive flow, we only focus on the transport of $n_s$ species. 

There is a duality in the DBO decomposition, in which $U$ is a low-rank basis in the physical domain and analogously   $Y$ constitutes a low-rank basis in the species space. As we demonstrate, DBO closely approximates the dominant \emph{instantaneous} correlated structures in the species. To formalize this connection, we define the species \emph{two-point correlation} operator  and its approximation by  DBO as follows:
\begin{equation}\label{eq:corr_tp}
   \mathcal{C}(x,x',t) =  \Phi(x,t) \Phi(x',t) ^T \simeq U(x,t)\Sigma(t) Y(t)^T Y(t) \Sigma(t)^T U(x',t)^T =  U(x,t) C(t) U(x',t)^T,
\end{equation}
where we have used the orthonormality condition $Y(t)^TY(t)=I$. In the above equation, $\mathcal{C}(x,x',t) \in \mathbb{R}^{\infty \times \infty}$ is the species two-point correlation operator. In the discrete representation,  $\mathcal{C}(x,x',t)$  is a matrix of size $N$ by $N$, where  $N$ is the total number of grid points, and   $C(t) = \Sigma(t)\Sigma(t)^T \in \mathbb{R}^{r \times r}$ is the low-rank correlation matrix. This shows that $\Sigma(t)$ is a factorization of the symmetric positive matrix $C(t)$.  As we  demonstrate, the eigenvalues of  $C(t)$ approximate the $r$ largest eigenvalues of $\mathcal{C}(x,x',t)$.

\subsection{Variational Principle}
In this section we present a variational principle whose optimality conditions lead to  closed form evolution equations for the components of the DBO decomposition. 
% The evolution equations of low-rank approximation are derived from a variational equation. 
%where 
%\begin{equation}
%\mathcal{M} = \sum_{n=1}^{ns}\int_{D}^{}\mathcal{M}_n^2dx.
%\\
%\end{equation}
The variational principle is given by:
\begin{equation}\label{eq:VarPrin}
\mathcal{F}(\dot{U}(x,t), \dot{\Sigma}(t), \dot{Y}(t)) = \left \|  \frac{\partial (U(x,t)\Sigma (t)Y(t)^T )}{\partial t}  - \mathcal{M}( \Phi)  \right \|_{\mathcal{F}}^2,
\end{equation}
subject to the orthonormality conditions of $u_i(x,t)$ and  $y_i(t)$ modes given by Eqs.\ (\ref{eq:U-Ortho})-(\ref{eq:Y-Ortho}). 
In Eq.~(\ref{eq:VarPrin}),  $\dot{( ~ )} = \partial ( ~ )/\partial t$ and $\mathcal{M}(\Phi)$ is the right hand side of species transport equation: $\mathcal{M}(\Phi) =  -(v \cdot \nabla )\Phi +  \nabla \cdot ( \nabla \Phi \bs{\alpha})+ S(\Phi, \rho,  T) $.  The variational principle given in Eq.\ (\ref{eq:VarPrin}) seeks to minimize the difference between the right hand side of the species transport equation  and the time derivative of the DBO decomposition subject to the orthonormality constraints given by Eqs.\ (\ref{eq:U-Ortho})-(\ref{eq:Y-Ortho}).  The  control parameters are $\dot{U}$, $\dot{Y}$ and $\dot{\Sigma}$. The orthonormality constraints can be incorporated into the variational principle via Lagrange multipliers. To this end, we first take a time derivative of the orthonormality constraints. This results in:  
\begin{align}
\inner{\dot u_i(x,t)}{u_j(x,t)} +\inner{u_i(x,t)}{\dot u_j(x,t)}  &= 0, \label{eq:Uortho} \\
\dot y_i^T(t) y_j(t) + y_i^T(t)\dot y_j(t)  &= 0. \label{eq:Yortho}
\end{align}
%\end{subequations}
 We denote $\varphi_{ij}(t) = \inner{u_i(x,t)}{\dot u_j(x,t)}$ and $\theta_{ij}(t) = y_i^T(t)\dot y_j(t)$. From Eqs. (\ref{eq:Uortho}) and (\ref{eq:Yortho}), it is apparent that both $\varphi(t)\in \mathbb{R}^{r\times r}$ and $\theta(t) \in \mathbb{R}^{r\times r}$ are skew-symmetric matrices, \textit{i.e.} $\varphi^T(t) = -\varphi(t)$ and $\theta^T(t) = -\theta(t)$. By incorporating the orthonormality constraints into the variational principle via Lagrange multipliers, the following unconstrained optimization problem is obtained:
\begin{equation}\label{eq:g_1}
\begin{aligned}
\mathcal{G}(\dot{U}, \dot{\Sigma}, \dot{Y}, \lambda,\gamma)  = 
                                  \left \|  \frac{\partial (U(x,t)\Sigma (t)Y(t)^T )}{\partial t}  - \mathcal{M}(\Phi)  \right \|_{\mathcal{F}}^2 
                                 + \lambda_{ij}(\inner{{u}_{i}}{\dot{u}_{j}} - \varphi_{ij}) + \gamma_{ij}(y_{i}^{T} \dot{y}_{j} - \theta_{ij}),
\end{aligned}
\end{equation}
where $\lambda_{ij}$ and $\gamma_{ij}$, $i, j=1, . . . , r$ are the Lagrange multipliers. In simple words, the variational principle seeks to \emph{optimally} update the DBO components given the change in the right hand side of the species transport equation while preserving the orthonormality constraints.  In Appendix \ref{apnA}, derivations of closed form evolution equations for $U(x,t)$, $\Sigma(t)$ and $Y(t)$ are provided  from the first-order optimality conditions of the functional (\ref{eq:g_1}). The closed form evolution equations of $U(x,t)$, $\Sigma(t)$ and $Y(t)$ are as follows: 
\begin{align}
%  \frac{\partial {u}_i}{\partial t}  &=   (\mathcal{M}(\Phi)y_{j} - u_n\inner{u_n}{\mathcal{M}(\Phi)y_{j}})\Sigma^{-1}_{ji} + u_n\varphi_{ni}, \label{eq:dudt} \\
 \frac{\partial U}{\partial t}  &=   (\mathcal{M}(\Phi)Y - U\inner{U}{\mathcal{M}(\Phi)Y})\Sigma^{-1} + U\varphi, \label{eq:dudtgen} \\
 \frac{d \Sigma}{d t} &= \inner{U}{ \mathcal{M}(\Phi)Y}  - \varphi \Sigma-\Sigma \theta, \label{eq:dSdtgen} \\
 \frac{dY}{dt}  &= (I-Y Y^T) \inner{\mathcal{M}(\Phi)}{U}\Sigma^{-T} + Y\theta. \label{eq:dydtgen}
\end{align}
As we show in  Appendix \ref{apnB}, any skew-symmetric choice for matrices $\theta$ and $\varphi$ leads to an equivalent DBO decomposition. In this work, we consider the simplest choice of $\varphi=\theta=0$. This choice lends itself to a simple interpretation: the updates  of the low-rank subspaces 
($\dot{U}$ and $\dot{Y}$) are orthogonal to the current subspace, \textit{i.e.}, $\inner{U}{\dot{U}}=0$ and $Y^T\dot{Y} =0$. This choice is referred to as \emph{dynamically orthogonal} condition and it was also used for the evolution of time dependent basis for the application of computing sensitivities \cite{Babaee_PRSA,DCB20} or stochastic reduced order modeling \cite{SL09,B19,PB20}.    We now replace $\mathcal{M}(\Phi)$ with the right hand side of the species transport equation (Eq.\ (\ref{eq:ADR})) and use the DBO low-rank decomposition for species, \textit{i.e.} $\Phi \cong U\Sigma Y^T$:
\begin{equation}
    \mathcal{M}(\Phi) = -(v \cdot \nabla) U \Sigma Y^T + \nabla \cdot (\nabla U \Sigma Y^T \bm{\alpha}) + S(U \Sigma Y^T, \rho,T).
\end{equation}
The projection of $\mathcal{M}(\Phi)$ onto the space spanned by the species modes is obtained by multiplying the above equation from right by matrix $Y$. This results in:
\begin{equation}\label{eq:MY}
\mathcal{M}(\Phi)Y = -(v \cdot \nabla) U \Sigma + \nabla \cdot (\nabla U \Sigma \bm{\alpha}_Y) + S(U \Sigma Y^T, \rho,T)Y,
\end{equation}
where $\mathcal{M}(\Phi)Y \in \mathbb{R}^{\infty\times r }$, and $\bm{\alpha}_Y = Y^T \bm{\alpha} Y \in \mathbb{R}^{r \times r}$ is the low rank diffusion matrix and for cases where $\bm{\alpha}$ is a function of $x$ and $t$, $\bm{\alpha}_Y$ is computed at each physical point $x$ and time $t$.
Similarly, the projection of $\mathcal{M}(\Phi)$ onto the spatial modes is obtained by taking the inner product of $\mathcal{M}(\Phi)$ with $U$ from right:
\begin{equation}\label{eq:MU}
\inner{\mathcal{M}(\Phi)}{U} = -Y\Sigma^T \inner{(v \cdot \nabla) U}{U}  + \inner{\nabla \cdot (\nabla U \Sigma Y^T  \bm{\alpha})}{U} + \inner{S(U \Sigma Y^T, \rho,T)}{U},
\end{equation}
where we have used $ \inner{(v \cdot \nabla) U\Sigma Y^T}{U}= Y\Sigma^T \inner{(v \cdot \nabla) U}{U}$. 
Substituting Eqs.\ (\ref{eq:MY}) and (\ref{eq:MU}) into Eqs.\ (\ref{eq:dudtgen}-\ref{eq:dydtgen}) yields:
\begin{align}
\mbox{Evolution of spatial modes (PDE):} \quad \frac{\partial U}{\partial t}  &=  \underset{\perp U}{\prod } \big[-(v \cdot \nabla) U + \nabla \cdot (\nabla U \Sigma \bm{\alpha}_Y)\Sigma^{-1} +  S Y\Sigma^{-1}\big], \label{eq:dudt} \\
\mbox{Reduced order model (ODE):} \quad \frac{d \Sigma}{d t} &= -\inner{U}{(v \cdot \nabla) U}\Sigma + \inner{U}{\nabla \cdot (\nabla U \Sigma \bm{\alpha}_Y)} + \inner{U}{SY}, \label{eq:dSdt} \\
\mbox{Evolution of species modes (ODE):} \quad \frac{dY}{dt}  &= \underset{\perp Y}{\prod } \big[  \inner{\nabla \cdot (\nabla U \Sigma Y^T\bm{\alpha}}{U} + \inner{S}{U}\big]\Sigma^{-T}, \label{eq:dydt}
\end{align}
where for $f(x) \in \mathbb{R}^{\infty \times 1}$ and $z \in \mathbb{R}^{n_s \times 1}$:
\begin{equation*}
    \underset{\perp U}{\prod } f  = f - U\inner{U}{f} \quad \quad \mbox{and} \quad \quad   \underset{\perp Y}{\prod } z  = z - YY^Tz,
\end{equation*}
are the orthogonal projections onto the complement of $U$ and $Y$, respectively.   In the above equations $S=S(U\Sigma Y^T,\rho,T)$. We make the following observations about Eqs. (\ref{eq:dudt}-\ref{eq:dydt}):

\begin{enumerate}
    \item The DBO decomposition has an on-the-fly built-in adaptivity to  ``chase"  the low-rank subspace that the species belong to.  This can be realized by noting that the  right hand side of the evolution equation of $U$, for example, is equal to the orthogonal projection of $R_U=\mathcal{M}(\Phi)Y\Sigma^{-1}$  onto the complement of $U$. This means that if $R_U$ is in the span of $U$, then  $R_U$ can be exactly expressed as a linear combination of $U$: $R_U = U C_U$ for some $C_U \in \mathbb{R}^{r \times r}$. In that case, $\dot{U}  = UC_U - U \inner{U}{UC_U} =   UC_U - U\inner{U}{U}C_U  = 0$, where we have used $\inner{U}{U}=I$. However, when $R_U$ is not in the span of $U$, the basis \emph{optimally} evolves to follow  the right hand side. An analogous mechanism  exists for the evolution of $Y$.
    \item The contribution of the convective terms to the evolution of $Y$ is zero. This is because the convective term $\inner{(v \cdot \nabla) U \Sigma Y^T}{U} =Y C_Y$ for  $C_Y=\Sigma^T \inner{(v \cdot \nabla) U  }{U} \in \mathbb{R}^{r \times r}$ is in the span of $Y$. Therefore, the projection of $Y C_Y$ onto the complement of $Y$ is zero. Similarly,  for the special case of space independent and equal diffusivity coefficient, \textit{i.e.} $\alpha_1=\alpha_2=\dots=\alpha_{n_s}$, the contribution of diffusive term to $\dot{Y}$ is also zero.  
    \item  Eq.~(\ref{eq:dSdt}) is a reduced order model (ROM) obtained by the projection of the full-dimensional species transport equation onto spatial and species modes. 
\end{enumerate}

The approximation error is computed as the difference between the  DBO reconstruction versus the full-rank solution of the species. We use a relative error to measure the accuracy of DBO decomposition. The relative error is defined as:
\begin{equation}
    \mathcal{E}(t) =\big  \| \Phi(x,t) - U(x,t) \Sigma(t) Y(t)^T \big \|_{\mathcal{F}}\big /\big  \| \Phi(x,t) \big \|_{\mathcal{F}}, 
\end{equation} 
where $\Phi(x,t)$ represent the solution of the full-dimensional species transport equation.

 There are two main distinctions between the stochastic DBO, which was recently introduced in \cite{PB20}, and Eq.~(\ref{eq:DBO}) that we highlight here: (i) Eq.~(\ref{eq:DBO}) approximates  deterministic fields (i.e., multiple species), whereas the stochastic DBO decomposes a random field, in which the  $y_i$ components are infinite-dimensional random processes. (ii)  In the stochastic DBO  a Reynolds decomposition of the stochastic field is considered, \textit{i.e.} the stochastic field is decomposed to the mean and fluctuations and the stochastic DBO decomposition is considered only for the approximation of the fluctuations. On the other hand, Eq. (\ref{eq:DBO}) does not consider a separate mean field. This simplifies the DBO evolution equations significantly. Moreover, in this work, it is shown that the evolution equations for the DBO components  minimize the variational principle  given by Eq.~(\ref{eq:g_1}). The variational principle can facilitate further adjustments to the DBO decomposition in a rigorous and unambiguous  manner. 

\subsection{Computational Cost}
The main computational advantage of using DBO is that the transport PDE is solved only for $r$ spatial modes (Eq. (\ref{eq:MU})) as opposed to $n_s$ species in the full transport equation. The computational cost of evolving  $\Sigma$ and $Y$ is negligible as they are governed by low-rank ordinary differential equations (ODEs). Moreover, in the DBO decomposition, the species are stored in the \emph{compressed form}, \textit{i.e.}, matrices $U$, $\Sigma$ and $Y$ are kept in the memory as opposed to their multiplication $U\Sigma Y^T$, \emph{i.e.,} the \emph{decompressed form}.  The memory storage requirement is dominated by $U$ as $\Sigma$ and $Y$ are low-rank matrices and their storage cost is negligible. Therefore,  in comparison to the full species transport equation, this results in the memory compression ratio of $n_s/r$. As we show below, if care is taken, it is possible to evolve the DBO components and not store the $\infty \times n_s$ species quasimatrix  in the decompressed at any point. Evidently, the same  compression ratio is gained when storing the time-resolved DBO solution to the disk, in which $U$, $\Sigma$ and $Y$ are written to the disk and any species can be reconstructed from the DBO decomposition.  The memory and I/O savings are very important for the future high performance computing architecture, where power restrictions impose stringent limitations on memory and I/O usage \cite{AM_DOE_14}.     In this section we discuss  how each term in the right hand side of Eqs.  (\ref{eq:MU}-\ref{eq:dydt}) can be computed. For brevity, we drop the dependence of $S$ on $\rho$ and $T$.
\begin{itemize}
    \item $\nabla \cdot (\nabla U \Sigma \bm{\alpha}_Y)$: This term can be written as 
$\nabla \cdot (\nabla U \Sigma \bm{\alpha}_Y) = \nabla \cdot (\nabla \tilde{U}) \in \mathbb{R}^{\infty \times r}$, where $\tilde{U}=UD$  and the low-rank matrix $D=\Sigma \bm{\alpha}_Y \in \mathbb{R}^{r \times r}$  can be computed point wise. This shows that computing this term requires computing Laplacian of $r$ field variables. This term is then utilized in Eqs. (\ref{eq:dudt}) and (\ref{eq:dSdt}).
\item $S(U\Sigma Y^T)$: This term can be computed in a loop over all grid points and when the species concentration at  point $x^*$ is required it can be computed as: $S(\Phi(x^*,t)) = S(U(x^*,t)\Sigma(t)Y(t)^T)$, where $U(x^*,t)\Sigma(t)Y(t)^T \in \mathbb{R}^{1\times n_s}$ is the vector of species concentration that  is utilized in  detailed kinetics. Again, the matrix $S(U\Sigma Y^T) \in \mathbb{R}^{\infty \times n_s}$ never have to be stored as only the projection of $S$ onto $Y$ and $U$, \textit{i.e.}, $SY \in \mathbb{R}^{\infty \times r}$ and $\inner{S}{U} \in \mathbb{R}^{n_s \times r}$  are needed in the DBO equations. Both  $\inner{S}{U}$ and $SY$ need to be stored and they  can be computed as a running summation and they can be stored in the same loop where $S$ is calculated. 
\item $\nabla \cdot (\nabla U \Sigma Y^T\bm{\alpha})$: Similar to the  computation of $S$, this term can be computed for each point $x^*$. This requires the computation of the  Laplacian of $n_s$ field variables. However,  only the projection of this term onto $U$  ($\inner{\nabla \cdot (\nabla U \Sigma Y^T\bm{\alpha})}{U}$) appears in the DBO equations, and this term can be computed as a running summation in the same loop over grid points. 
\end{itemize}

% For the case where the diffusivity coefficient is not a function space, i.e., $\bs{\alpha}=\bs{\alpha}(t)$ 

\subsection{Canonical Representation}
The DBO spatial and species modes can be ranked based on ``energy" in the second norm sense.   The ranking can be achieved by performing  a singular value decomposition (SVD) of $\Sigma(t)$:
\begin{equation}
    \Sigma(t) = R_U(t)\tilde{\Sigma}(t)R_Y(t),
\end{equation}
where $\tilde{\Sigma}(t)$ is a diagonal matrix that contains the ranked singular values: $\tilde{\sigma}_1(t) > \tilde{\sigma}_2(t) >...> \tilde{\sigma}_r(t)$ and $R_U(t)$ and $R_Y(t)$ are orthonormal matrices that can be used to rotate $U$ and $Y$ as follows:
\begin{subequations}\label{eq:USigmaY}
\begin{align}
\tilde{U}(x,t) &= U(x,t)R_U(t), \label{RankedU}\\
\tilde{Y}(t) &= Y(t)R_Y(t). \label{RankedY}
\end{align}
\end{subequations}
The components $\{\tilde{U}(x,t), \tilde{\Sigma}(t),\tilde{Y}(t)\}$ represent the DBO decomposition in the canonical form. We note that the DBO in the canonical form and the form that is computed are equivalent: $U\Sigma Y^T = \tilde{U}\tilde{\Sigma} \tilde{Y}^T$. See Appendix \ref{apnB} for a more details on equivalent decompositions. In any demonstration figures in this paper, the components of the DBO decomposition are shown in the canonical form. 

\subsection{Static vs. Time-Dependent Manifolds}
\label{subsection:DBO-PCA}
In this section, we draw contrast between time-dependent manifolds extracted by  DBO  and the static manifolds extracted from PCA. To this end, we write the DBO and PCA decompositions  in the matrix form: 
\begin{subequations}
\begin{align}
\mbox{PCA decomposition:} \quad \Phi(x,t) &\simeq  U_{PCA}(x,t)Y_{PCA}^T,\label{eq:PCA} \\ 
\mbox{DBO decomposition:} \quad \Phi(x,t) &\simeq  U(x,t)\Sigma (t)Y(t)^T\label{eq:DBO-M}.
\end{align}
\end{subequations}
In the PCA decomposition, $Y_{PCA} \in \mathbb{R}^{n_s \times r}$ represents the static manifold. In PCA,  the full-dimensional spatiotemporal species  data  is required  in the form of: $A=[\Phi^T(x^*,t^*)] \in \mathbb{R}^{n_s \times n}$, where $x^*$ and $t^*$ are selected points and times, respectively and $n$ is the total number of space-time points.   The data on species  $\Phi(x,t)$ is obtained by either a high-fidelity simulation, \textit{e.g.} DNS or from experiment. The matrix $Y_{PCA}$ consists of the first $r$ eigenvetors of the correlation matrix: $\overline{C}=A A^T \in \mathbb{R}^{n_s \times n_s}$ associated with the $r$ largest eigenvalues of $C$.  An evolution equation for the principal component $U_{PCA}(x,t)$ is obtained by plugging the decomposition, given in Eq.\ (\ref{eq:PCA}), into  Eq.\ (\ref{eq:ADR}) and perform the Galerkin projection onto $Y_{PCA}$.  
There are several key differences between DBO and PCA decompositions: (i) the PCA decomposition relies on high-fidelity data, \textit{i.e.} one must  commit to  preforming a high-fidelity simulation of a canonical problem with all species, or conducting an experiment and storing the high-dimensional data. That may be a prohibitively expensive undertaking for problems with a very large number of species both  in terms of computation/experiment and storage requirements of the space-time resolved matrix $A$. On the other hand, the DBO decomposition does not require data: it extracts the low-dimensional structure directly from the species transport equation. In that sense, DBO eliminates the potentially expensive and \emph{offline} step of high-fidelity data generation  that is required in the PCA workflow. We note that the PCA workflow is similar to many other common data-driven reduction techniques such as proper orthogonal decomposition (POD) \cite{A91,ABGA15} and dynamic mode decomposition (DMD) \cite{S10,KBBJ16,LKB18,VEKK19}. (ii) Reliance of the PCA approximation to high-fidelity training data means that $Y_{PCA}$ may not be a good low-rank representative of the full  composition space when used for different operating conditions, \textit{e.g.} boundary conditions, geometry, Mach and Reynolds numbers.   This potential limitation does not exist for the DBO decomposition, because DBO  is solved for the problem at hand. (iii) In the DBO decomposition, the low-rank matrix $Y(t)$ is a time-dependent manifold, whereas in PCA, $Y_{PCA}$ is static. This allows the DBO decomposition to instantaneously adapt to  changes in $\Phi(x,t)$.  More specifically,  $Y_{PCA}$ is a low-dimensional  manifold in a \emph{time-averaged} sense. This can be realized by inspecting the continuous analogue of the eigen-decomposition of the correlation matrix $\overline{C}$ that is formed in PCA: $\overline{C}_{ij} =\frac{1}{T} \int_T \int_D \phi_i(x,t) \phi_j(x,t) dx dt$.
As we demonstrate in our results, the DBO decomposition closely approximates the eigen-decomposition of the \emph{instantaneous} correlation matrix:  $\hat{C}_{ij}(t) = \int_D \phi_i(x,t) \phi_j(x,t) dx$. 

We  compare the performance of DBO with the  reduction based on instantaneous principal component analysis (I-PCA). The I-PCA components can be computed in a data-driven manner  by computing SVD of the instantaneous matrix of full species: $\Phi(x,t)=\hat{U}(x,t)\hat{\Sigma}(t)\hat{Y}^T(t)$, where $\hat{( \sim )}$ denotes the components of I-PCA,  $\hat{U}(x,t)=\{\hat{u}_1(x,t), \hat{u}_2(x,t), \dots, \hat{u}_{n_s}(x,t)\}$ is the quasimatrix of left singular functions, $\hat{\Sigma}(t) = \mbox{diag}(\hat{\sigma}_1(t),\hat{\sigma}_2(t), \dots, \hat{\sigma}_{n_s}(t))$ is the diagonal matrix of singular values and $\hat{Y}(t) = \{\hat{y}_1(t),\hat{y}_2(t), \dots, \hat{y}_{n_s}(t) \}$ is the matrix of right singular vectors.  It is also straightforward to show that the I-PCA spatial and species modes are the eigenfunctions and eigenvectors of the two-point correlation operator (Eq.~(\ref{eq:corr_tp})) and instantaneous correlation matrix, respectively, as shown below:     
\begin{equation}
    \int_D \mathcal{C}(x,x',t)\hat{u}_i(x',t)dx' = \hat{\sigma}_i^2(t)\hat{u}_i(x,t) \quad \mbox{and}  \quad \hat{C}(t) \hat{y}_i(t) = \hat{\sigma}_i^2(t)\hat{y}_i(t), \quad  i=1,2, \dots, n_s.
\end{equation}
A rank-$r$ truncated I-PCA represents the best  approximation that any rank $r$ decomposition in the form of Eq.~(\ref{eq:DBO}) can achieve at any given time $t$ and therefore, comparison of DBO components in the canonical form with I-PCA  shows how closely DBO is approximating the optimal low-rank decomposition. 

\section{Demonstration Cases} \label{Demonstration Cases}
\subsection{1D Passive Transport}
In this section, we demonstrate the application of DBO in solving passive transport equation with many species. The passive species are governed by Eq.~(\ref{eq:ADR}) for the case of $S(\Phi,\rho,T) = 0$. We consider $n_s=1000$  species with different diffusivities and initial conditions. The solution of Eq.~(\ref{eq:ADR}) is considered with $v(x,t)$  obtained from the solution of Burger's equation with the presence of shocks. We  compare the solution obtained by the  DBO reduction against the same-rank I-PCA reduction.

% We consider passive scalar equation given by:

% \begin{equation}\label{PST}
%     \frac{\partial \phi_i(x,t)}{\partial t} + v(x,t) \frac{\partial \phi_i(x,t)}{\partial x} = \alpha_i \frac{\partial ^{2}\phi_i(x,t)}{\partial x^{2}}
% \end{equation}
% where $\Phi(x,t)$ is the scalar, $v(x,t)$ is the velocity of the incompressible flow and $\alpha_i$ is the diffusivity of the $i$_th scalar in the fluid.

\begin{figure}[!tb]
 \begin{subfigmatrix}{2}
  \subfigure[Passive species]{\includegraphics[]{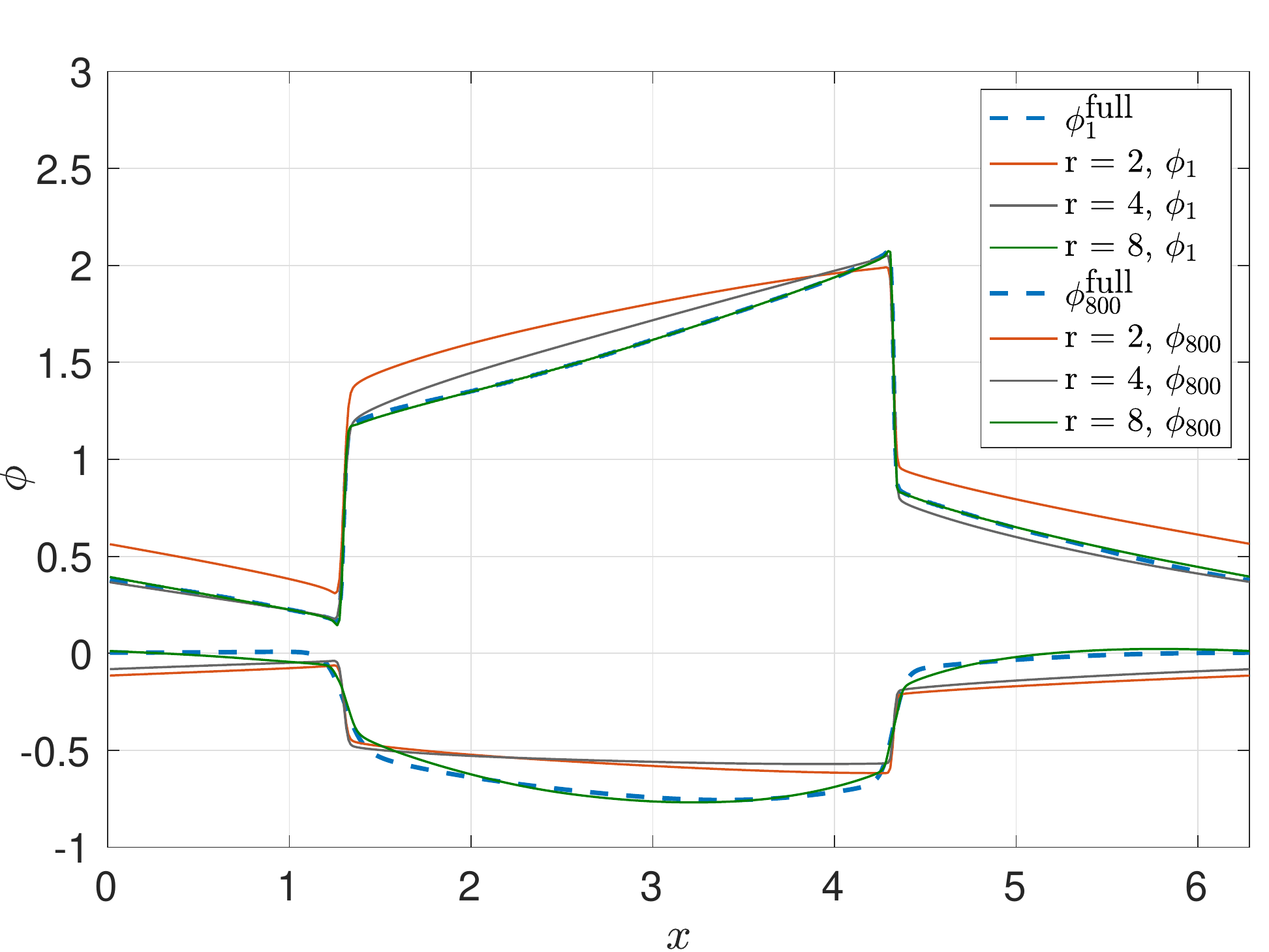}}
  \subfigure[Spatial modes]{\includegraphics{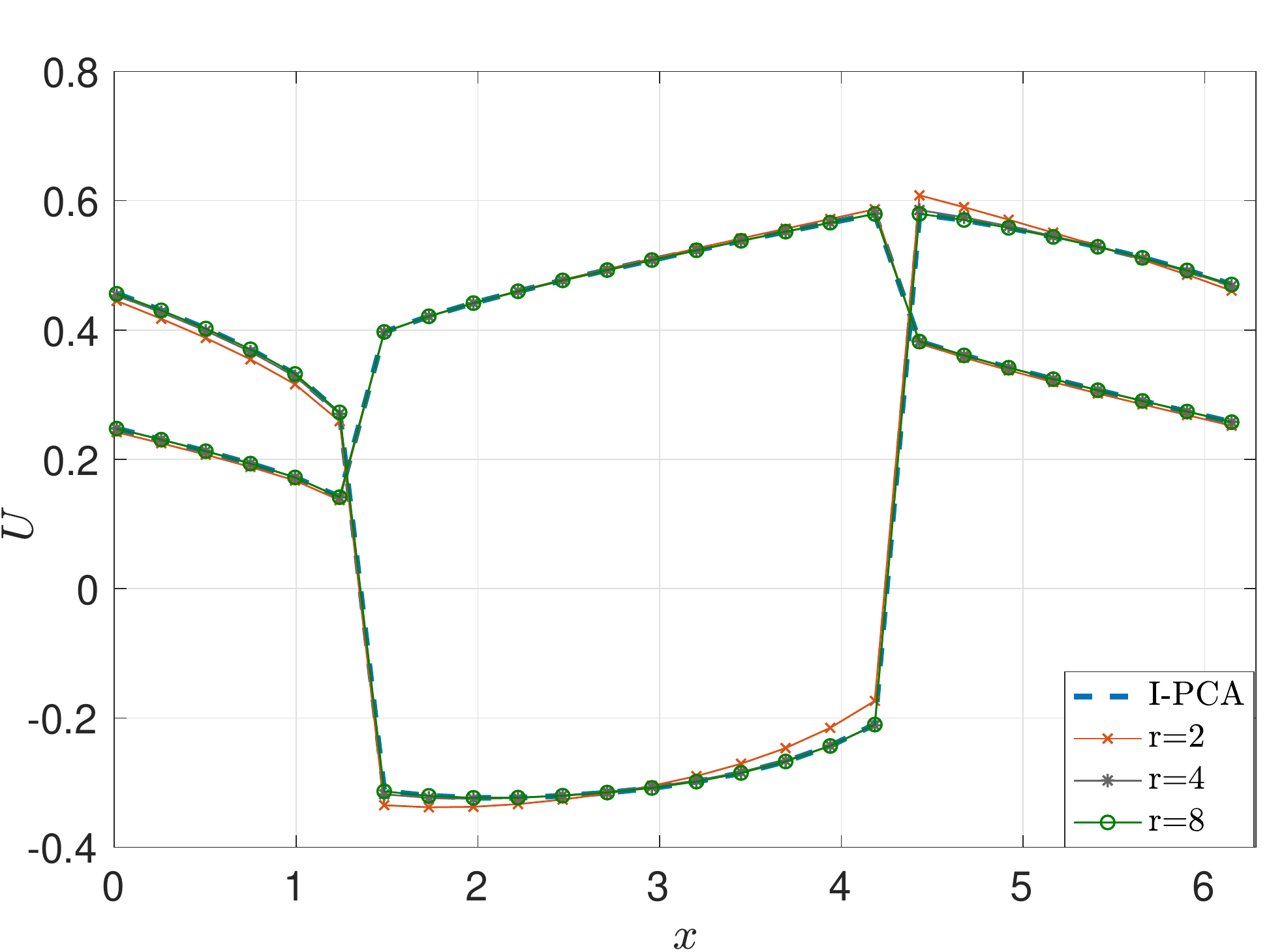}}
 \end{subfigmatrix} \caption{DBO of passive species transported in 1D Burger's equation. (a) Comparison of solutions of $\phi_1$ and $\phi_{800}$ obtained from DBO using reduction sizes of $r=2,4$ and $8$ with full dimensional passive species solutions over the physical domain at $t=4$. (b) Comparison of the first two dominant spatial modes using reduction sizes of $r=2,4$ and $8$ at $t=4$ with the optimal spatial modes obtained from I-PCA. }
 \label{u-phi}
\end{figure}

We consider the Burgers' equation governed by: 
\begin{equation}\label{Burgers}
    \frac{\partial v}{\partial t} + v \frac{\partial v}{\partial x} = \nu \frac{\partial ^{2}v}{\partial x^{2}} , \ \ \  x\in [0,2\pi], \ \ \ \text{and} \ \ \ t\in [0,t_f],
\end{equation}
where $v$ is the velocity of the flow and $\nu$ is the viscosity of the fluid which is assumed to be $0.01$. The Burgers' equation is  solved with the initial condition of $v(x,0) = 0.5(\exp (\cos x)-1.5)\sin(x+2\pi\times0.37)$
and the initial condition of the passive species transport equation is assumed to be:
  $ \phi_i(x,0) = \sum_{n=1}^{n_s}\frac{\zeta_i^{(n)}}{n^b} \sin (\frac{n \pi x}{L})$,
where $\zeta_i^{(n)}$ are  chosen from  an independent normal distribution  and $b$ is the rate of decay of the spectrum and is considered to be $2$. The diffusivity of the $i$th species is considered to be $\alpha_i = 0.01/\sqrt i, i=1, \dots, n_s$ and the length of the physical domain is  $L=2\pi$. The reason that each species has a different concentration is because of different initial condition and different diffusivity coefficients.  The DBO spatial modes, governed by Eq. (\ref{eq:dudt}), are solved using spectral Fourier method with $N = 512$ Fourier modes. The fourth-order Runge-Kutta scheme is used for the time integration of the DBO Eqs. (\ref{eq:dudt}-\ref{eq:dydt}) with $\Delta t = 1/256$ and up to the final time of $t_f = 4$ time units. 
In Figure~\ref{u-phi}(a), solutions of two different passive species ($\phi_1$ and $\phi_{800}$) using  reduction sizes of $r=2,4$ and $8$ at $t=4$ are shown.  The presence of the traveling shock in $v$ manifests itself with a sharp  change in the passive species.   As it can be seen, even for $r=2$, \textit{i.e.}, a drastic reduction, the location of the shock is predicted correctly and as  reduction rank increases from $r=2$ to $r=8$ the passive species obtained by DBO converge to the true profile obtained by directly solving the full-dimensional passive species and the location and amplitude of the shock are correctly captured.  The accuracy of DBO is also compared against the I-PCA. As Fig.\ \ref{u-phi}(b) represents, the first two I-PCA spatial modes ($\hat{u}_1(x,t)$ and $\hat{u}_2(x,t)$), which are the first two dominant eigenfunctions of $\mathcal{C}(x,x',t)$, match relatively well with the  DBO spatial modes ($\tilde{u}_1(x,t)$ and $\tilde{u}_2(x,t)$) for $r = 2$, and as the reduction size increases to  $r=4$ and $r=8$ the agreement between the first two DBO modes and those of the  I-PCA improves.

\begin{figure}[!tb]
 \begin{subfigmatrix}{2}
  \subfigure[Error]{\includegraphics[]{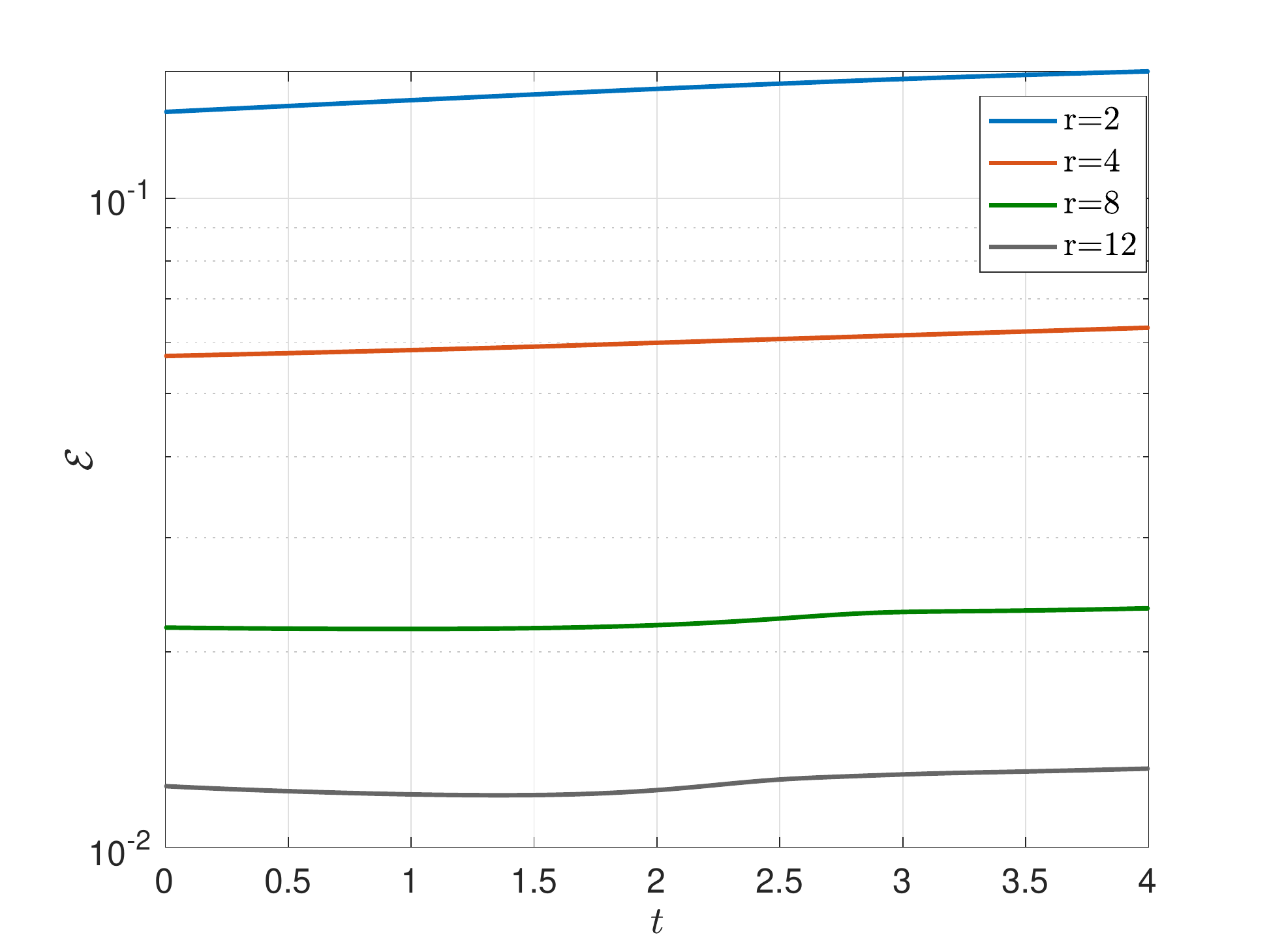}}
  \subfigure[Singular values]{\includegraphics[]{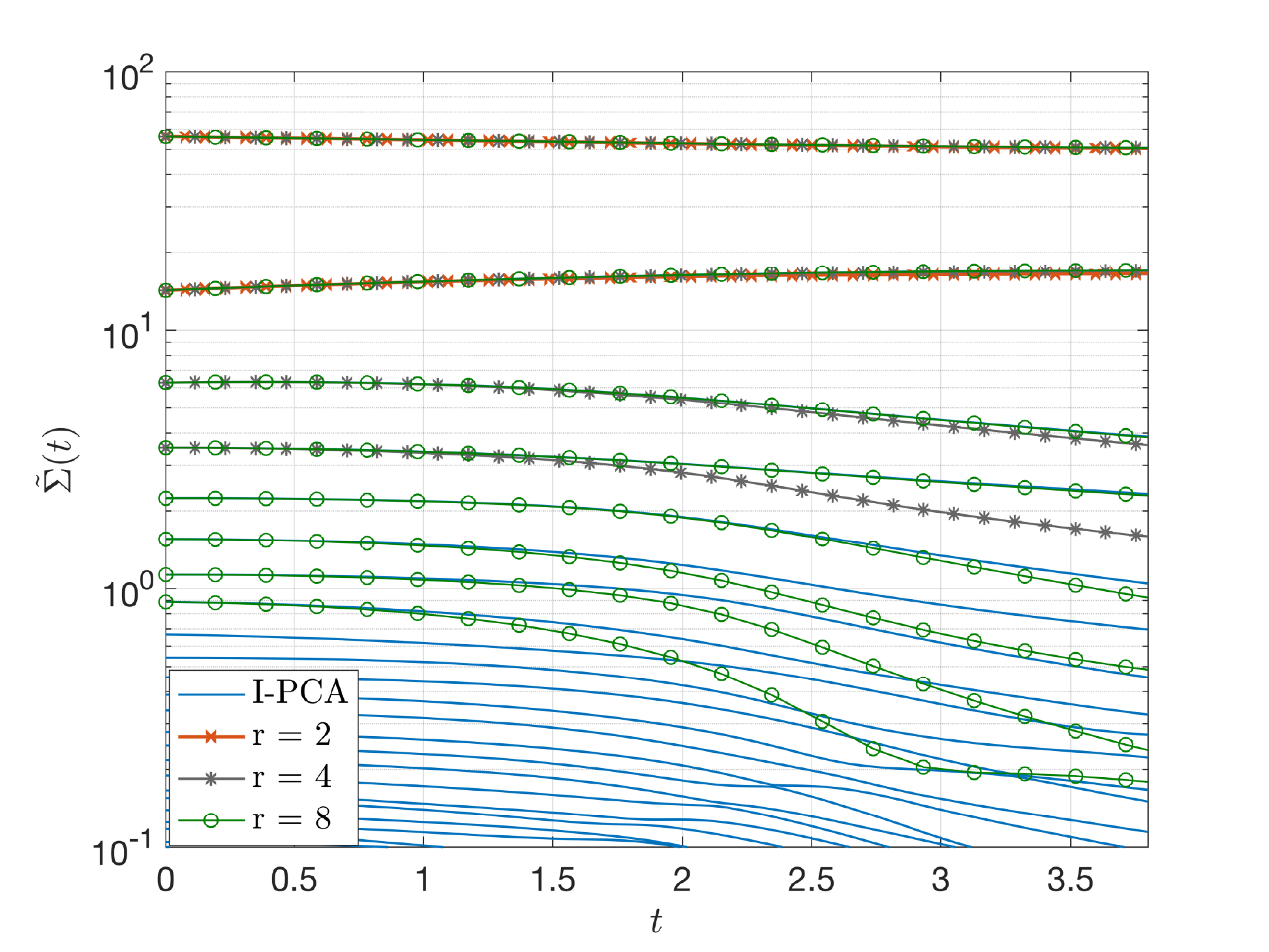}}
 \end{subfigmatrix} \caption{DBO of passive species transported in 1D Burger's equation: (a) Evolution of solution error using reduction sizes of $r=2,4$,$8$ and $12$. (b) Evolution of singular values of I-PCA and DBO using reduction sizes of $r=2,4$ and $8$.}
 \label{Modes-Error}
\end{figure}

In Fig.\ \ref{Modes-Error}(a) the evolution of the error using different reduction sizes are shown. The error improves about one order of magnitude by increasing the reduction size from $r=2$ to $r=12$. In addition, the evolution of singular values extracted from I-PCA and DBO solutions are presented in Fig. \ref{Modes-Error}(b). It is observed that DBO can  capture the largest singular values  with a better accuracy. However, the accuracy of DBO degrades for capturing lower singular values. The difference between singular values of I-PCA and those of DBO is the result of the lost interactions of DBO modes with the unresolved modes. The unresolved error affects  the modes with smaller singular values more intensely. Ultimately, the DBO can be augmented with an adaptive strategy to add/remove modes based on a criterion. Similar strategies have been adopted in the past. See for example \cite{Babaee:2017aa}.      For a discussion of the error in ROM based on time-dependent basis, see  \cite{DCB20}.

\subsection{Incompressible Reactive Flow}
We investigate the performance of DBO  for an incompressible reactive flow, in which velocity is not affected by reactions and   the reactive source term is a function of $\Phi$ only, \textit{i.e.}, $S(\Phi)$. To this end, we use DBO to solve a biochemically reactive flow,  in which the coagulation cascade process in a Newtonian fluid is modeled  by Anand \textit{et al.} \cite{anand2005model}. In this model, the coagulation cascade process is simulated by solving for $n_s=$23 species. The  velocity field is obtained by solving  the incompressible  Navier-Stokes equations:

\begin{figure}[!tb]
\begin{center}
 \includegraphics[width=.8\textwidth]{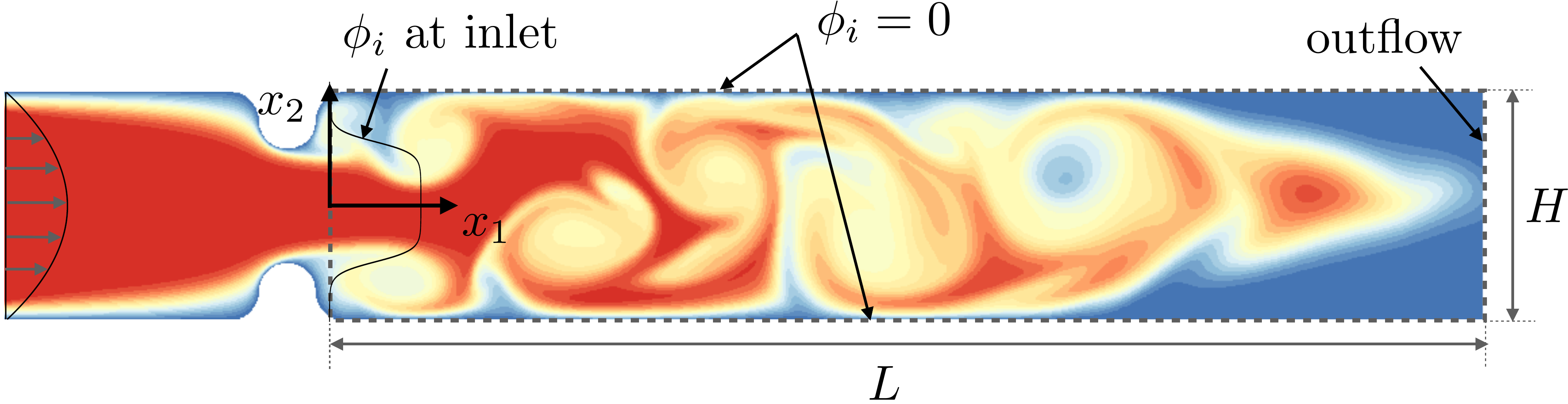}
 \caption{2D incompressible turbulent reactive flow: Schematic  of the computational domain and boundary conditions.}
 \label{Jetschem}
 \end{center}
\end{figure}

\begin{equation}\label{Navier-Stookes}
\begin{aligned}
\frac{\partial v}{\partial t} + (v \cdot \nabla) v &= -  \nabla p + \frac{1}{Re} \nabla^2 v, \\
\nabla \cdot v &= 0,
\end{aligned}
\end{equation}
 where $v = (v_x, v_y)$ is the velocity vector field, $p$ is the pressure field,  and $Re$ is the Reynolds number. The involved species, their corresponding Schmidt numbers, as well as source terms of each reactant can be found in \cite{2015transport}.

The schematic of the problem is shown in Figure\ \ref{Jetschem}.  Height and length of the jet are $H=2$ and $L=10$, correspondingly and  the Reynolds number based on reference length of half the height ($H/2$) and the kinematic viscosity $\nu$  is $Re=\overline{w}H/2\nu=1000$.   The species boundary condition at the inlet is $\phi_i(0,x_2,t) = 1/2\big(\tanh{(x_2+H/2)/\delta} -\tanh{(x_2-H/2)/\delta}  \big)$ for all species, where $\delta = 0.1$.  For the spatial discretization, spectral/hp element method is used with $N_e =4008$ quadrilateral elements and polynomial order of $5$. For more  details of the spectral/hp element method  see \cite{KS05,Babaee:2013ab,Babaee:2013aa}. The fourth-order Runge-Kutta scheme is used for the time integration with $\Delta t = 5 \times 10^{-4}$ and the system is  solved till $t_f= 5$ units of time. The velocity field at each time step is then interpolated on a Cartesian spectral element mesh in the rectangular domain shown by dashed line in Figure~\ref{Jetschem}. The DBO  equations are solved in the rectangular domain with 251 elements in the horizontal direction and 76 elements in the vertical direction. The  spectral polynomial of order 5 is used in each direction.

\begin{figure}[!tb]
\begin{center}
 \includegraphics[width=1\textwidth]{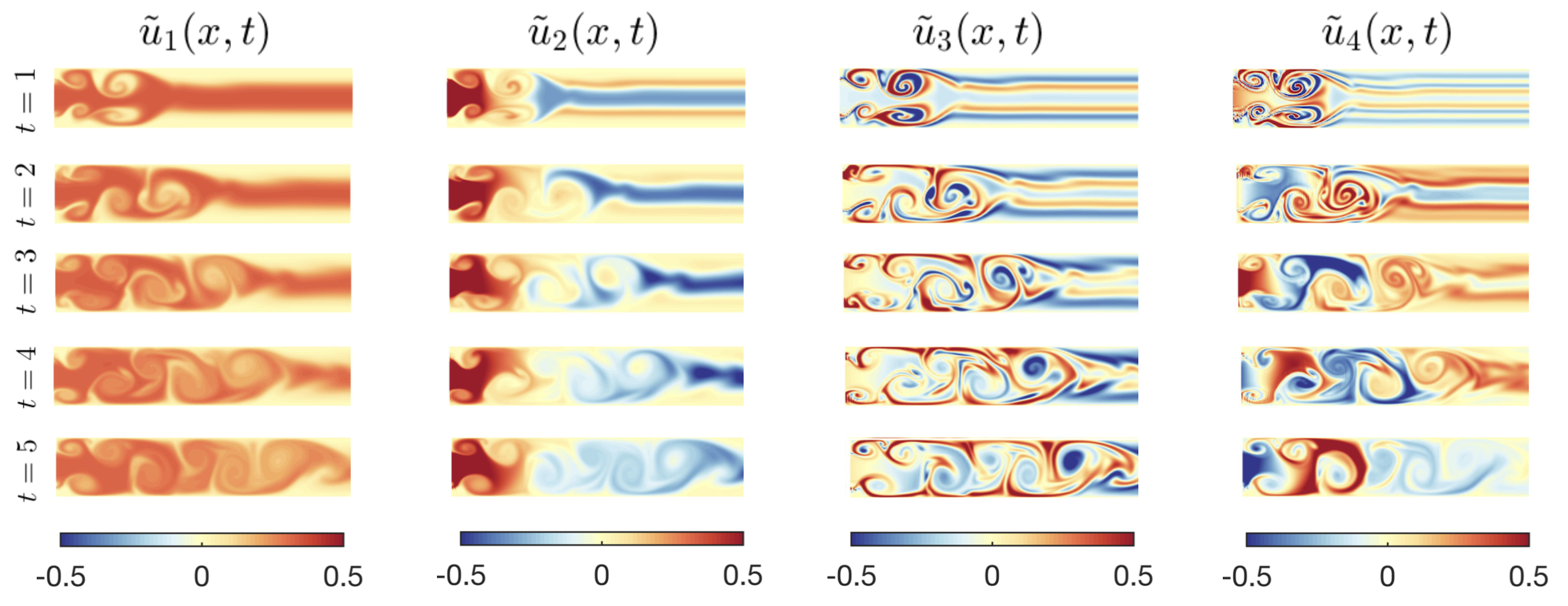}
 \caption{2D incompressible turbulent reactive flow. The  four most dominant DBO spatial modes in different times.}
 \label{Modes-ADR}
 \end{center}
\end{figure}

%%% Discussion of results
  We solve the problem using $r=5$ reduction size. Figure~\ref{Modes-ADR} shows the  first $4$ dominant spatial modes at 5 different instants. The first mode, associated with the largest singular value, is  positive in all $x$ and $t$.  This mode is the most energetic mode and it roughly approximates the orthonormalized mean of the species, \textit{i.e.}, $\tilde{u}_1(x,t) \simeq \overline{\phi}(x,t)/\| \overline{\phi}(x,t)\|$, where  $\overline{\phi}(x,t) = 1/n_s \sum_{i=1}^{n_s} \phi_i(x,t)$. This occurs when the mean of the species is in fact the most energetic mode.  The higher modes change sign in the domain and capture finer structures. Also it is clear that the modes are advecting with the flow from left to right. 
  \begin{figure}[!h]
  \centering
 \includegraphics[width=.8\textwidth]{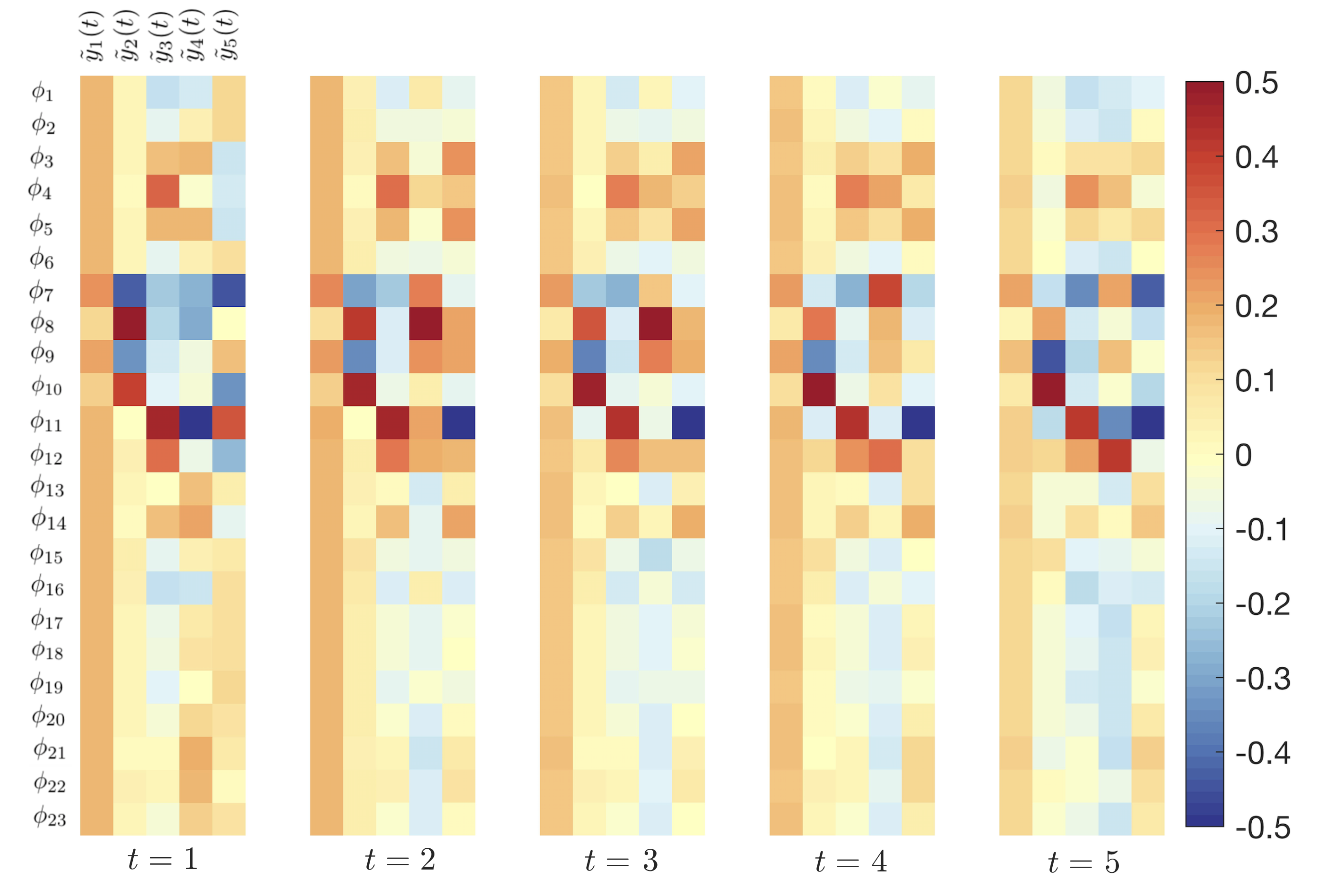}
%   \subfigure[Difference between IPCA and DBO hidden species
%   modes]{\includegraphics{Plots/Ydiff.png}}
 \caption{Time-dependent low-dimensional manifold of 2D incompressible turbulent reactive flow:   dominant species modes in different times in the form of color-coded matrices.}
 \label{y-ADR}
\end{figure}
\begin{figure}[!h]
\begin{center}
 \includegraphics[width=.9\textwidth]{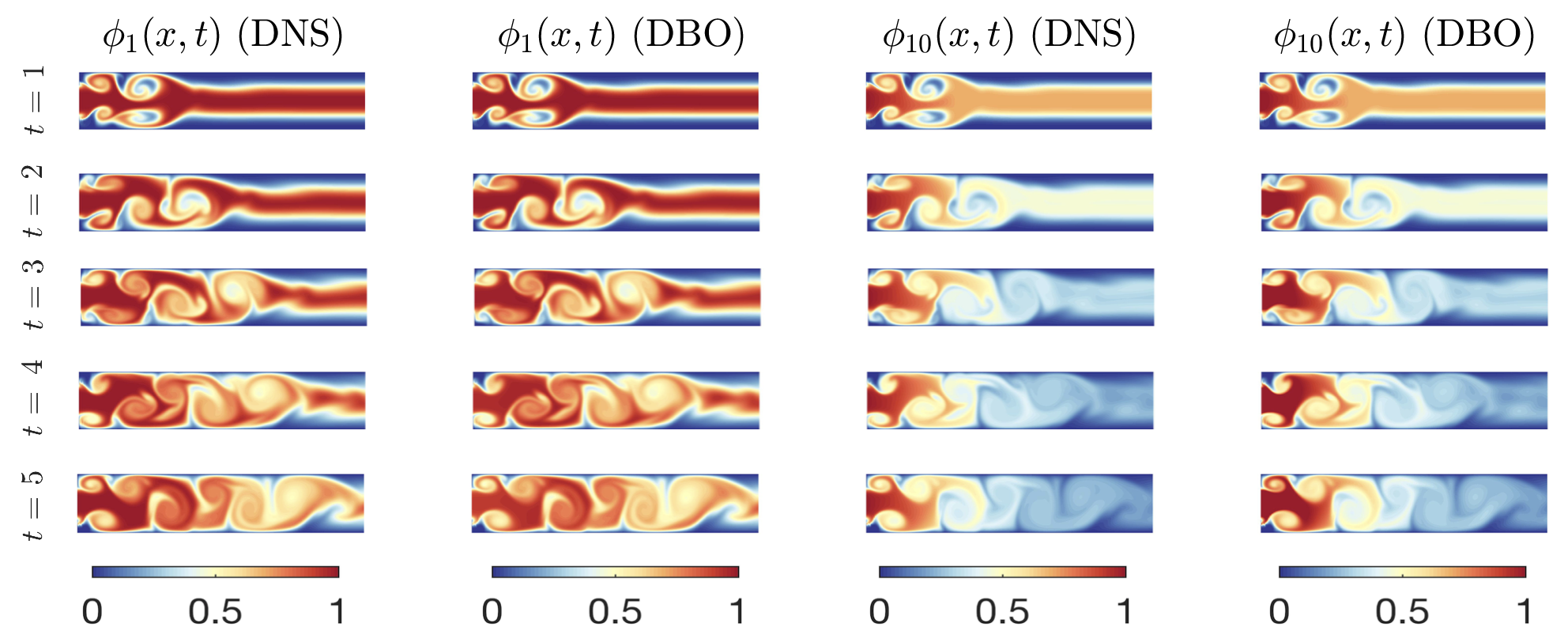}
 \caption{ 2D incompressible turbulent reactive flow: species concentration  of first and tenth species in different times obtained from DNS (full-dimensional) and DBO using $r=5$. }
 \label{Phi-Turb-ADR}
 \end{center}
\end{figure}

Figure \ref{y-ADR} displays the species matrix $\tilde{Y}(t)$ in different times. Similar to the spatial modes, the first mode associated with the most energetic direction is always positive. This mode changes slowly  with time. On the other hand the higher modes have positive and negative contribution of each species and they change faster with time. Each component of vector $\tilde{y}_i(t)$ should be interpreted as the instantaneous contribution of the corresponding species. The species labels are shown in Figure \ref{y-ADR}.    Unlike conventional reduction schemes such as skeletal reductions, DBO does not eliminate any species or reactions. Instead, a time-dependent weighted   contributions of all species are considered in the low-rank decomposition. For example, as it is clear from Figure~\ref{y-ADR}, $\phi_8$ and $\phi_{10}$ have a small component in $\tilde{y}_1(t)$, which vanishes with time. On the other hand, $\phi_{7}$  and $\phi_9$ have a larger footprint in all $\tilde{y}_i$ modes.    

\begin{figure}
 \begin{subfigmatrix}{2}
 \subfigure[Singular values]{\includegraphics[]{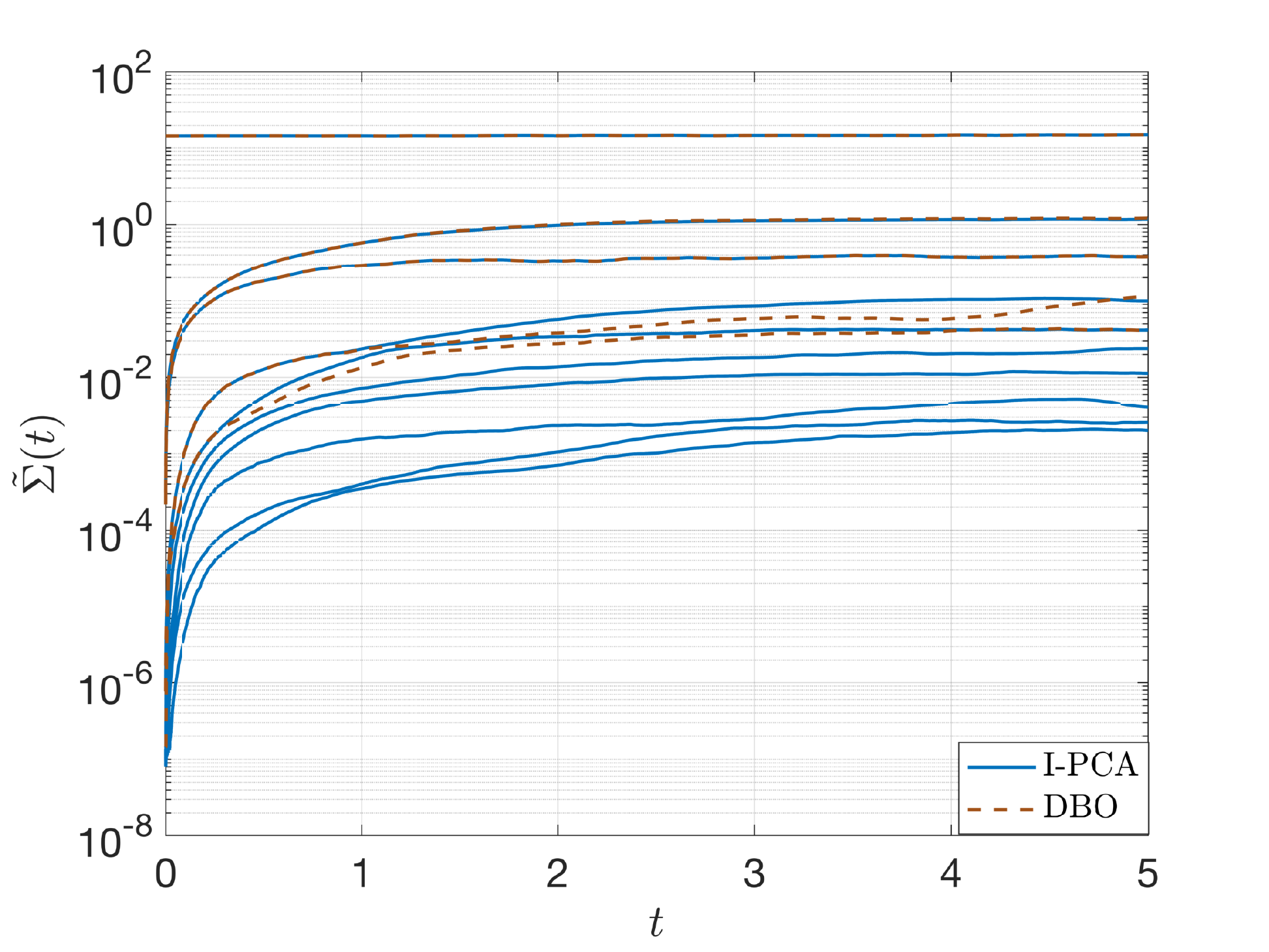}}
  \subfigure[Solution error]{\includegraphics[]{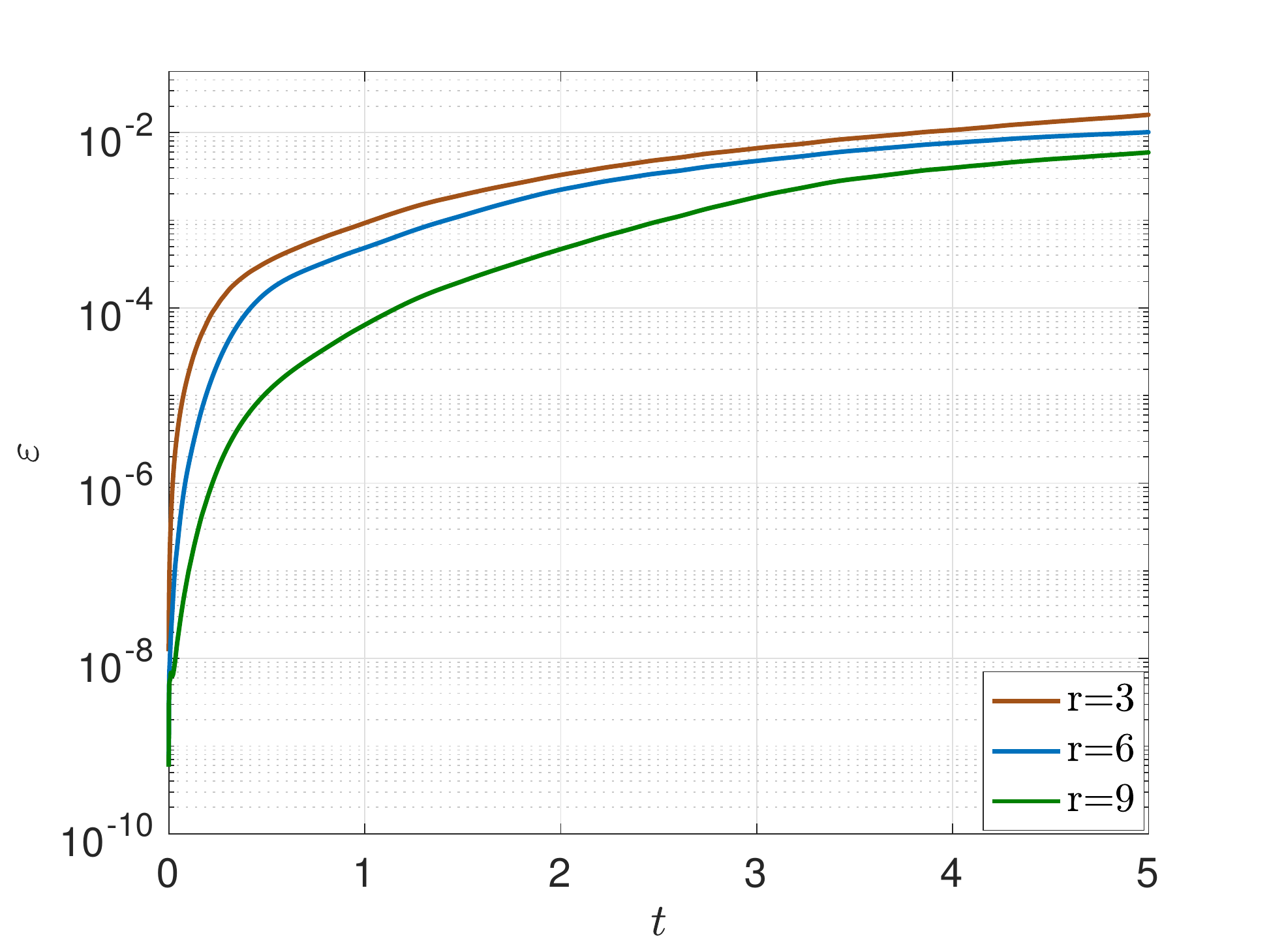}}
 \end{subfigmatrix} \caption{2D incompressible turbulent reactive flow. (a) Evolution of eigenvalues obtained from DBO solutions using $r=5$ reduction size and I-PCA solutions. (b) Evolution of solution error using reduction sizes of $r= 3,6$ and $9$.} 
 \label{Turb-Sigma-Error}
\end{figure}

Figure~\ref{Phi-Turb-ADR} shows several snapshots  of first and tenth species concentration  obtained from direct numerical solution (DNS) of the full-dimensional species transport and DBO. We can observe that solutions obtained from DBO using only $r=5$ are very similar to DNS data at all times.  For further examining the accuracy of DBO, singular values extracted from DBO decomposition are compared with those of  the I-PCA decomposition. As presented in Figure~\ref{Turb-Sigma-Error}(a), the evolution of singular values extracted from DBO solutions closely match with I-PCA, and similar to the previous demonstration case the most dominant modes show better agreement with I-PCA. In Figure~\ref{Turb-Sigma-Error}(b), the approximation errors of DBO for $r=3,6$ and 9  versus time are shown.  This shows that the reduction error decreases by increasing 
$r$.

% \begin{comment}
% \begin{figure}
% \begin{center}
%  \includegraphics[width=.6\textwidth]{Plots/Turb-y.png}
%  \caption{ADR equations for a 2D incompressible turbulent flow in a jet with 25 chemical reactant species: magnitudes of the $4$ dominant hidden-species modes in different times in the form of color-coded matrices.}
%  \label{y-ADR}
%  \end{center}
% \end{figure}
% \end{comment}

\subsection{Compressible Reactive Flow}
The applicability of using DBO decomposition for simulating 2D compressible reactive Navier-Stokes is demonstrated in this section. For the mathematical description of compressible flows involving $n_s$ species, the primary transport variables are the density $\rho(x,t)$, velocity vector $v(x,t)$, pressure $p(x,t)$, total energy $E(x,t)$,
temperature $T(x,t)$, and species mass fractions $\Phi(x,t)$. The  reactive  compressible Navier-Stoke equations are given  by:

\begin{subequations}\label{eq:ReactiveComp_NS}
 \begin{align}
\frac{\partial \rho}{\partial t} + \frac{\partial \rho v_j}{\partial x_j} &= 0,\\
\frac{\partial \rho v_i}{\partial t} + \frac{\partial \rho v_i v_j}{\partial x_j} &= - \frac{\partial p}{\partial x_i} + \frac{\partial  \tau_{ij}}{\partial x_j}, \\
\frac{\partial \rho E}{\partial t} + \frac{\partial (\rho E v_j)}{\partial x_j} &= - \frac{\partial p v_j}{\partial x_j} + \frac{\partial ( \tau_{ij}v_i)}{\partial x_j} - \frac{\partial q_j} {\partial x_j} + W(\Phi, \rho,  T), \label{eq:energy_compressible}\\
\frac{\partial  \Phi_k}{\partial t} + v_j\frac{\partial \Phi_k }{\partial x_j} &=  -\frac{1}{\rho} \frac{\partial J^k_j}{\partial x_j} + S_k(\Phi, \rho,  T).
\end{align}
\end{subequations}
Here, the viscosity flux $\tau$, heat flux $q$, mass flux  $J$, and total energy are represented by: 
\begin{equation*}\label{tauandE}
\tau_{ij} = \frac{1}{Re} \Big(\frac{\partial v_i}{\partial x_j} + \frac{\partial v_j}{\partial x_i}- \frac{2}{3}\frac{\partial v_k}{\partial x_k} \delta_{ij}\Big),\ \ 
q_j = -\frac{1}{Ec.Pe}  \frac{\partial T}{\partial x_j}, \ \
J^k_j = -\alpha_k \frac{\partial \phi_k }{\partial x_j},
\end{equation*}
in dimensionless format where $e$ is the internal energy, $E = e + \frac{1}{2}v_iv_i$ is the total energy, $Ec = (\gamma -1)Ma^2$, $Pe = Re.Pr$, and $Ma$ are Eckert, Peclet, and Mach numbers, respectively. Moreover, $W=\sum_{k=1}^{n_s} \rho S_k \Delta h_{f,k}^0$ is the  heat release where $S_k$ is the dimensionless species source term and $\Delta h_{f,k}^0$ is the formation enthalpy of species $k$. We assume a perfect gas with the specific heat ratio $\gamma = c_p/c_v$ and $R =c_p-c_v$, where $R$ is the gas constant, $c_p$ and $c_v$ denote the specific heats at 
constant pressure and constant volume, respectively, and are assumed to be constants. Equal and constant diffusion
coefficients  and unity Schmidt ($Sc = \mu/D$) and Prandtl ($Pr = c_p \mu/\lambda$) numbers for all  species are assumed, where $D$ is the mass diffusivity and $\lambda$ is the heat conductivity. The viscosity and molecular diffusion coefficients can, in general, be temperature dependent but in this study, they are assumed to be constants. As the ability of DBO decomposition in handling different species with different diffusivities was demonstrated in previous sections, here we use a simple diffusion model. More complex simulations with realistic conditions will be the subject of future studies. 
\begin{figure}[tb]
\begin{center}
 \includegraphics[width=0.9\textwidth]{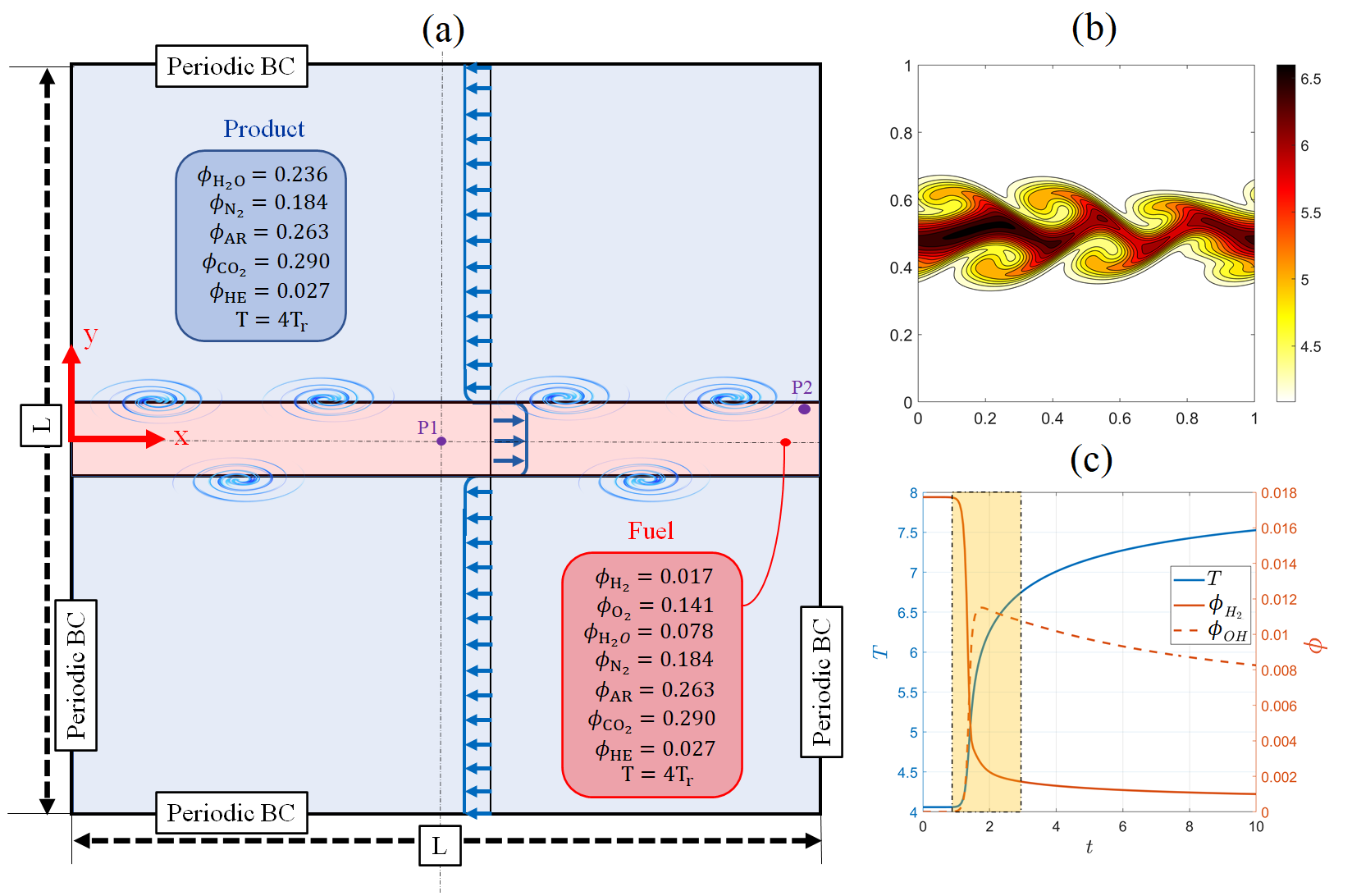}
\caption{Compressible reactive flow. (a) Computational domain and boundary conditions for compressible reactive Navier-Stokes simulation with $15$ species for hydrogen-air burning based on the model in Ref.~\cite{Keromnes13}, (b) contour of temperature at $t=3$, (c) dimensionless temperature as well as mass fractions of $H_2$ and $O_2$ in a zero-dimensional constant pressure reactor initialized with $T=4$ and mass fractions from the reference state. In 2D simulation the species and energy source terms are multiplies by a constant (0.01 here) to experience temperature inflection point in the middle of the domain at the time of vortex roll-ups.}
\label{fig:Comp_NS}
 \end{center}
\end{figure}

Simulations are conducted of a 2D temporally evolving jet transport of 15 species\footnote{$H,H_2,O,O_2,OH,H_2O,N_2,HO_2,H_2O_2,AR,CO,CO_2,OH^*,HE,HCO$} ($n_s =15$) associated with hydrogen-air burning based on the  model in Ref.~\cite{Keromnes13}. As is shown in  Fig.\ \ref{fig:Comp_NS}(a) our temporal layer consists of three parallel streams. The middle stream travels in opposite direction but with the same speed as the bottom and top streams. The transport variables are normalized with respect to $L_r$, $V_r$, $\rho_r$, and $T_r$ where $L_r$ is the size of domain in each direction and $V_r = \Delta V$ is the velocity difference across the layer. $\rho_r$ and $T_r$ are defined for a mixture of species in which $\phi^{m}_{H_2}=2 \phi^{m}_{O_2}=2\phi^{m}_{N_2}=2\phi^{m}_{AR} =2\phi^{m}_{CO_2} = 2\phi^{m}_{HE}=2/7$, under $T = 300K$ and $p = 1$ atm where $\Phi^{m}_k$ is the mole fraction of $k$th species. In our simulations, $Re = \rho_rV_rL_r/\mu = 10^4$ and $Ma = V_r/\sqrt{\gamma R T_r} = 0.5$. Periodic boundary condition are imposed on all four boundaries (Figure\ \ref{fig:Comp_NS}(a)) and initial temperature and pressure on the entire domain are set to $T = 4.0$ and $p=2.871$.  Species mass fractions are initialized as shown in  Figure\ \ref{fig:Comp_NS}(a). Figure\ \ref{fig:Comp_NS}(b) shows the contour of temperature at $t=3$.  For synchronizing hydrodynamics with combustion by equating ignition delay\footnote{Time of temperature inflection point $dT/dt|_{max}$ in the fuel side of the domain} and vortex roll-up time, the species source terms are multiplied by $0.01$. Figure\ \ref{fig:Comp_NS}(c) portrays the evolution of dimensionless temperature and mass fractions of $H_2$ and $O_2$ in a constant pressure ($p=2.871$) zero-dimensional reactor with initial temperature of $T = 4$ and initial mass fractions shown in the fuel side of Figure\ \ref{fig:Comp_NS}(a). It is clear in Figure\ \ref{fig:Comp_NS}(c) that the temperature inflection point happens at $t\approx 1.5$. The physical domain is discretized using the Fourier spectral method with 256 Fourier modes, and we used ode113~\cite{SR97} of MATLAB for time integration of DBO components.

\begin{figure}[!t]
\begin{center}
 \includegraphics[width=.95\textwidth]{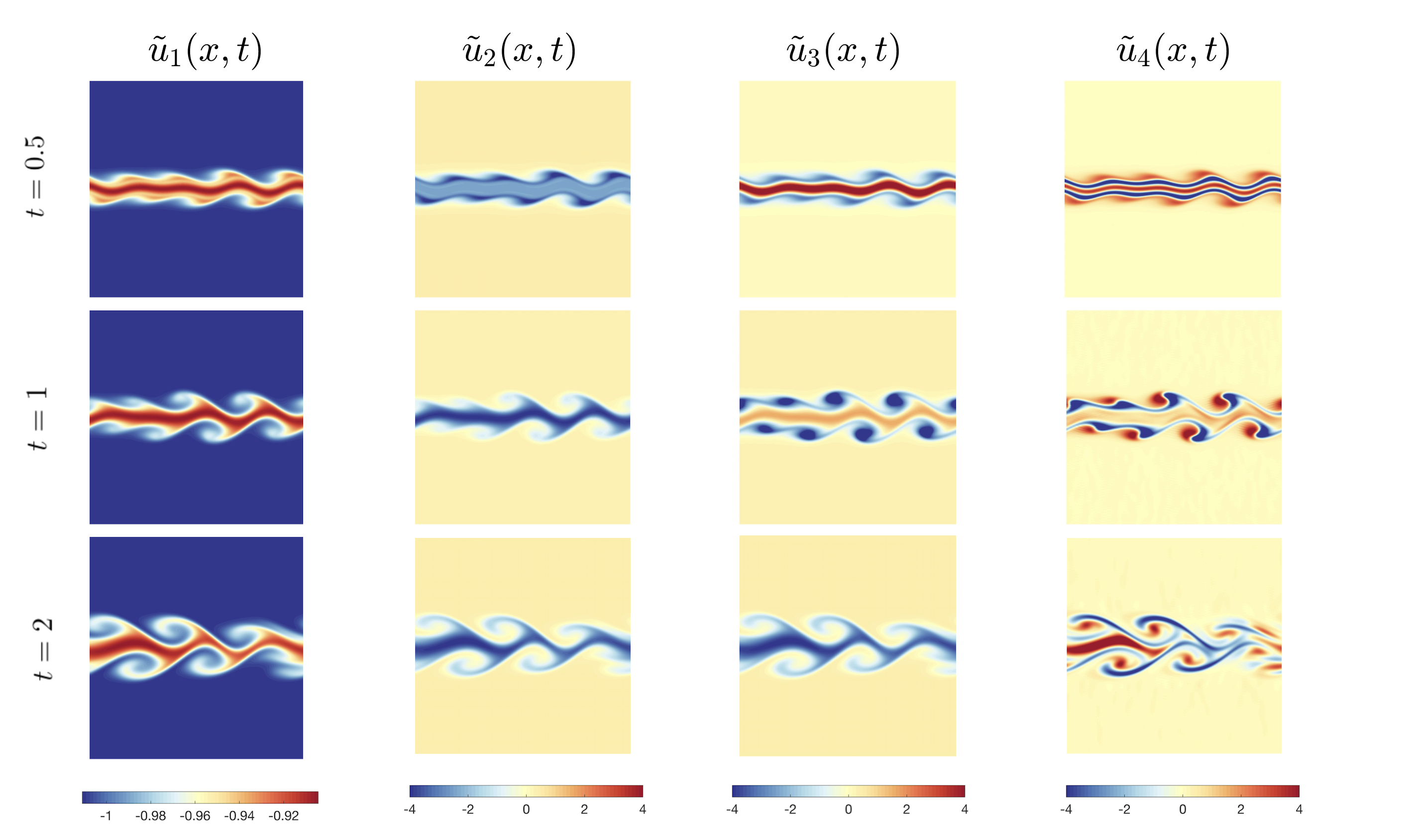}
 \caption{Compressible reactive flow. First four orthonormal DBO spatial modes in different time. }
 \label{fig:Phi-Compress}
 \end{center}
\end{figure}

DBO and PCA-ROM simulations are initialized at $t=1$ from the DNS data and evolved with $r$ modes till $t=3$, \textit{i.e.} within the orange rectangle in Fig.\ \ref{fig:Comp_NS}(c). PCA-ROM modes containing a linear combination of $n_s$ species mass fractions were extracted from the DNS data of all the spatial points of the domain and at 200 equally spaced time steps from $t=1$ to $t=3$. Here, the results of DBO and PCA-ROM are shown and compared against DNS within $\Delta t = 2$ after their initialization ($t \in [0,2]$). Figure\ \ref{fig:Phi-Compress} portrays the evolution of the vortexes and first four spatial modes of DBO ($\tilde{u}_i(x,t)$).

\begin{figure}[!h]
 \begin{subfigmatrix}{2}
 \subfigure[Singular values]{\includegraphics[]{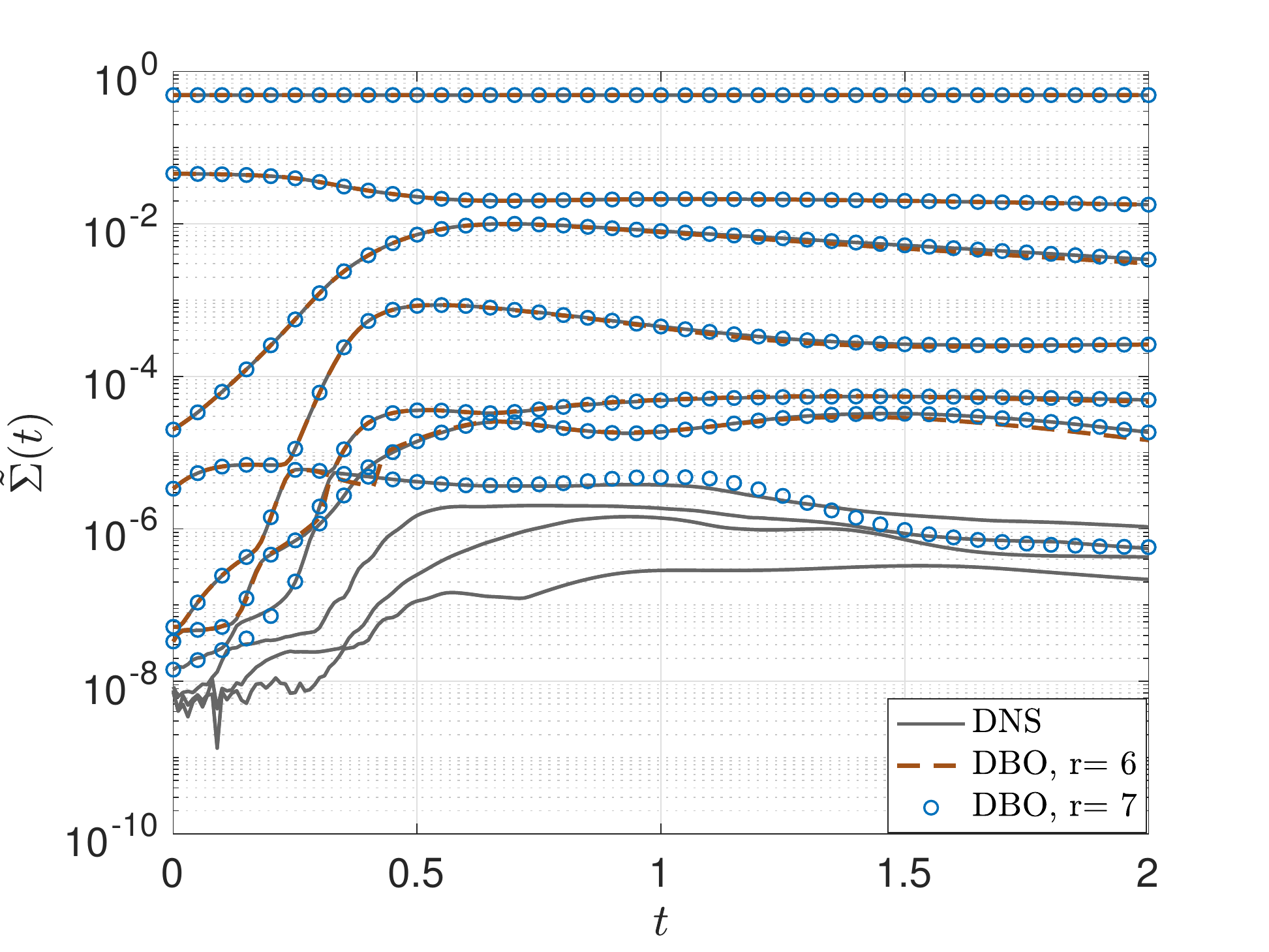}}
  \subfigure[Species error]{\includegraphics[]{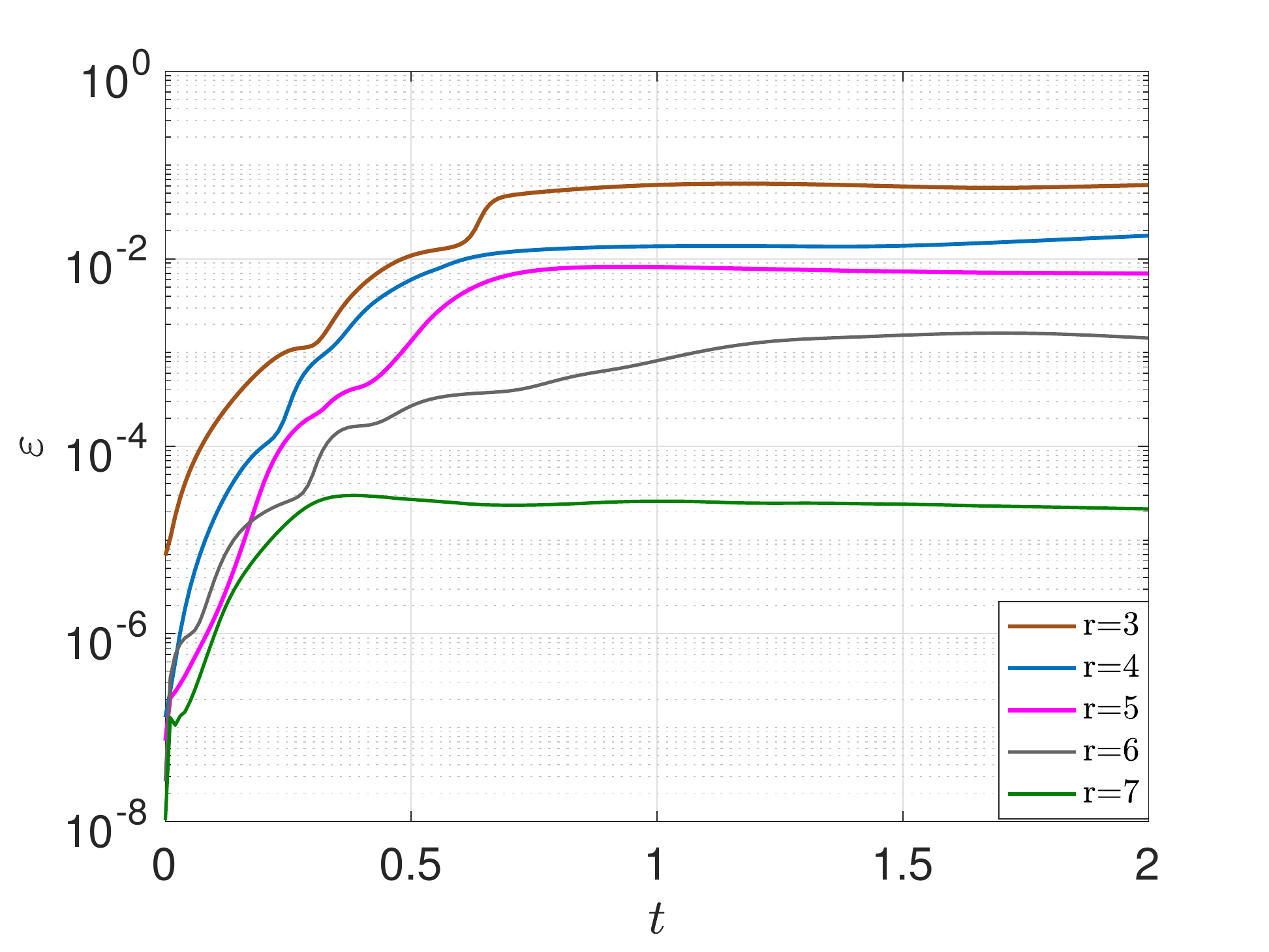}}
 \newline
 \subfigure[Temperature error]{\includegraphics[]{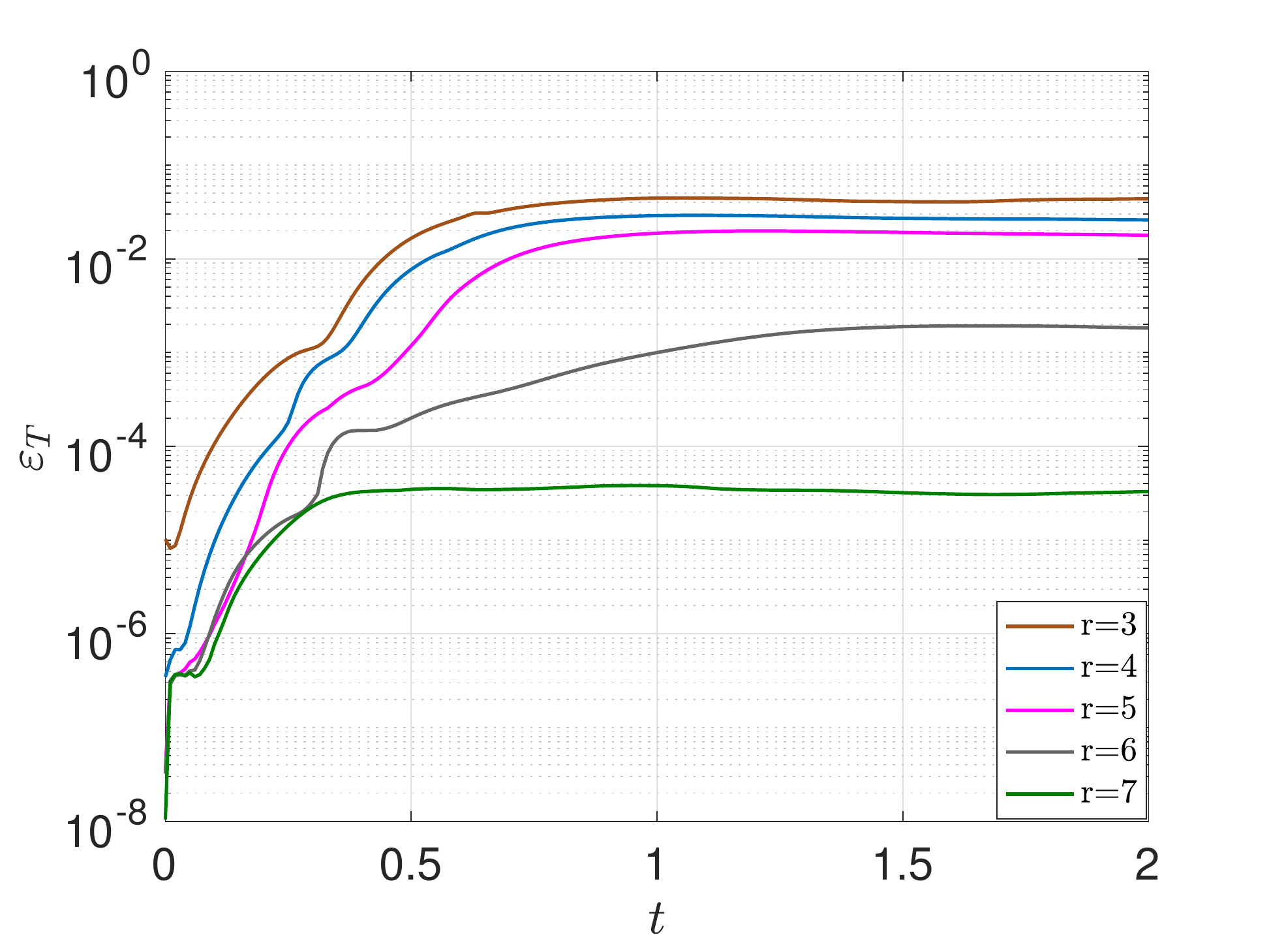}}
  \subfigure[Density error]{\includegraphics[]{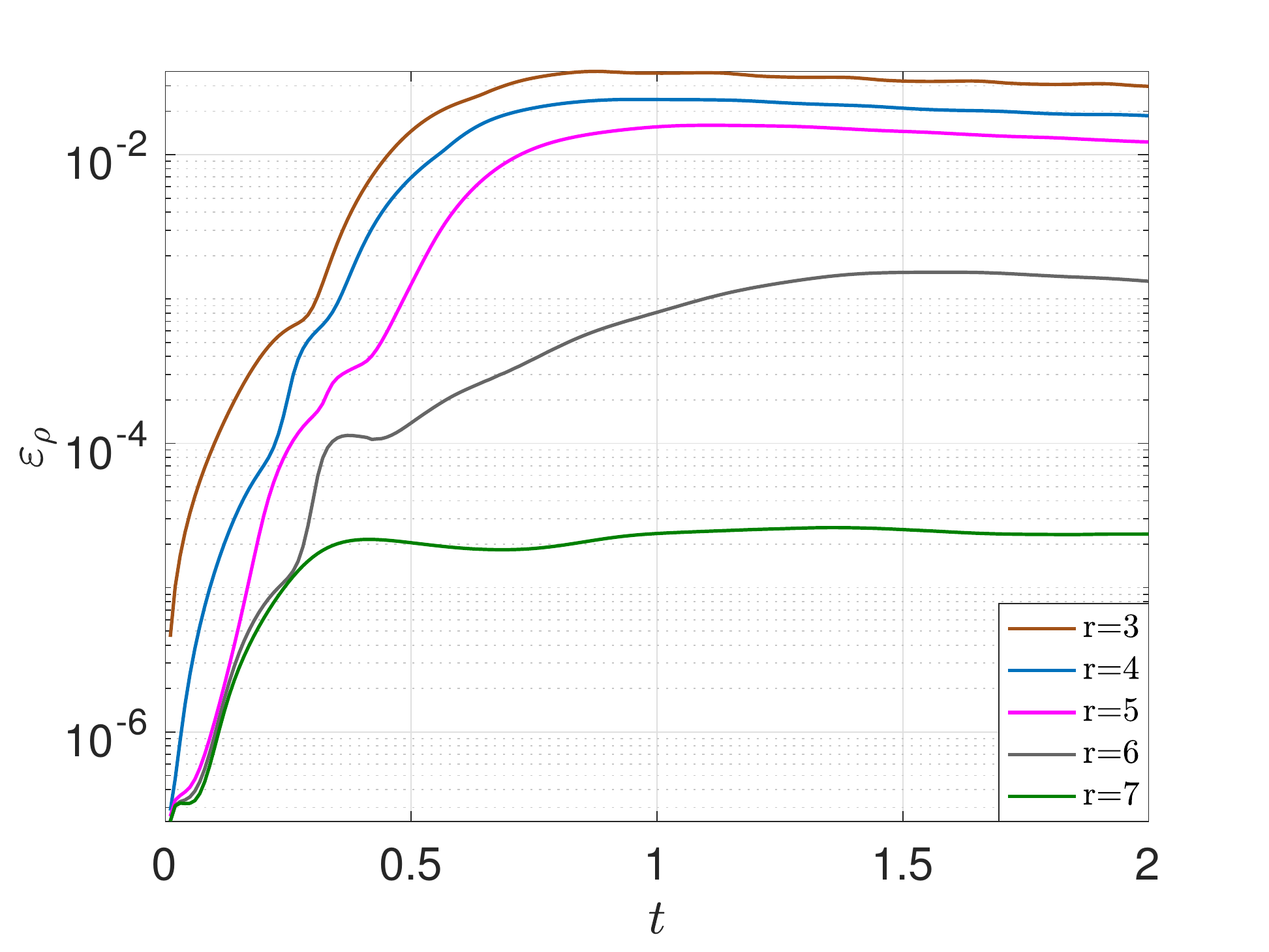}}
 \end{subfigmatrix} \caption{Compressible reactive flow. (a) Evolution of eigenvalues extracted from DNS and DBO  using $r=6$ and $7$, and (b) errors in  species mass fractions, (c) temperature, and (d) density using $r=3,4,5,6$ and $7$ in DBO simulations.}
 \label{fig:Ns-Error}
\end{figure}

Figure\ \ref{fig:Ns-Error}(a) shows the instantaneous singular values of   DBO decomposition for $r=6$ and $7$ as well as the I-PCA singular values. It is clear that the singular values of DBO very closely match with the most dominant  singular values of I-PCA. For the compressible reactive flow, the DBO approximation error affects $\rho$, $\rho v_i$ and $E$ due to two-way nonlinear coupling of species and these variables, since the heat source $W(\Phi,\rho,T)$ utilizes $\Phi \simeq U\Sigma Y^T$.    Figure\ \ref{fig:Ns-Error}(b),(c) and (d) demonstrate the instantaneous errors in estimating $\Phi, T,$ and $\rho$, respectively.  In general the error decreases with increasing $r$. %However, there are some instances in Fig.\ \ref{fig:Ns-Error}(b) in which the estimated $\Phi$ with $r=5$ is closer to DNS rather than the simulation with $r=6$ because of eigenvalue crossing. 

%%%%
% \begin{figure}[htb]
% \begin{center}
%  \includegraphics[width=.5\textwidth]{Plots/PCA-Temp.png}
%  \caption{Evolution of temperature in a constant pressure zero dimensional reactor initialized with $\bm{\Phi}$, $T$, and $P$ of the center point of the domain in $t=1$. The PCA modes are constructed based on data gathered from detailed 2D simulation from $t=1$ to $t=3$.}
%  \label{PCA-Temperature}
%  \end{center}
% \end{figure}

\begin{figure}[tb]
 \begin{subfigmatrix}{2}
 \subfigure[Temperature at P1]{\includegraphics[]{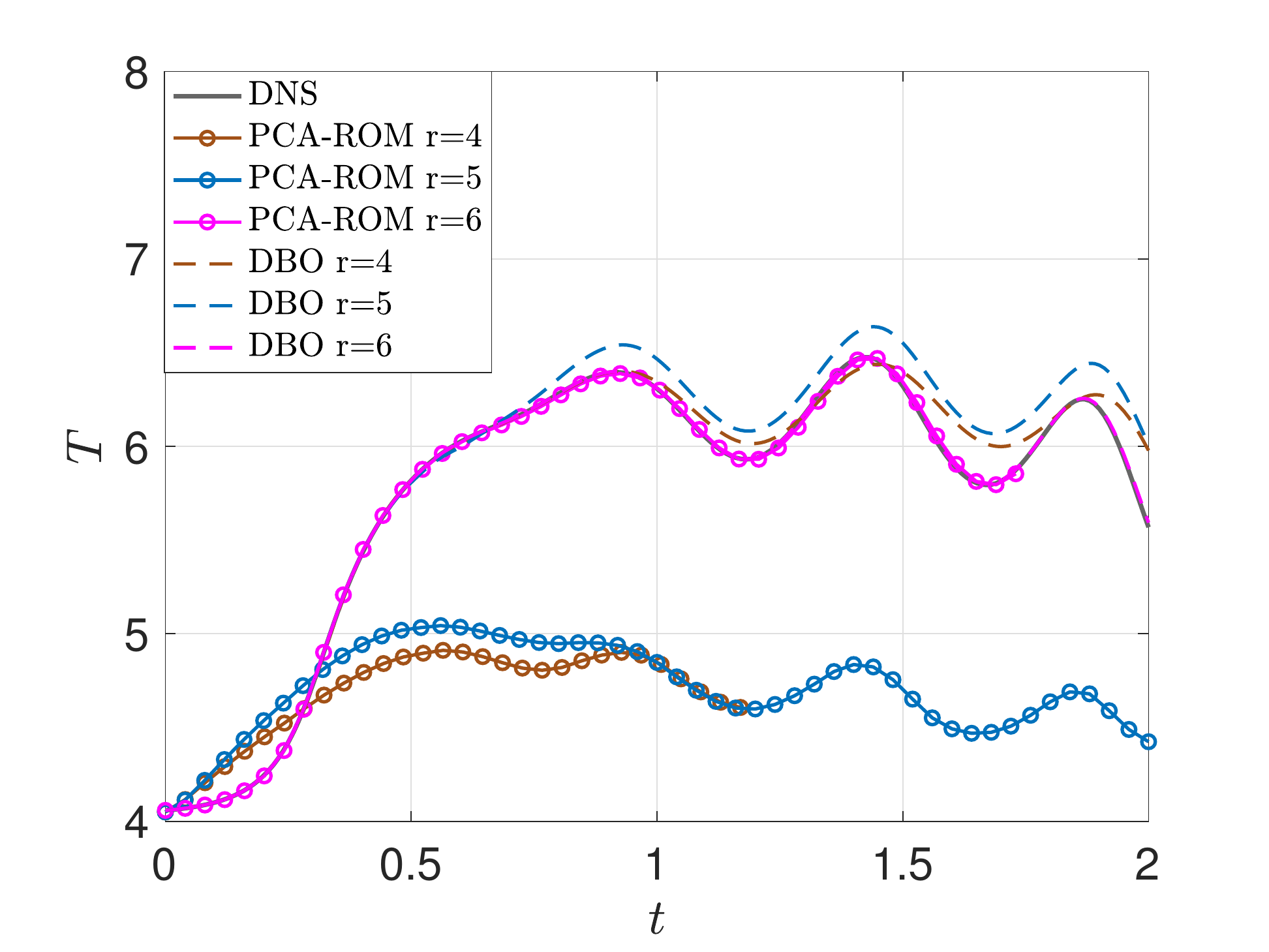}}
  \subfigure[Temperature at P2]{\includegraphics[]{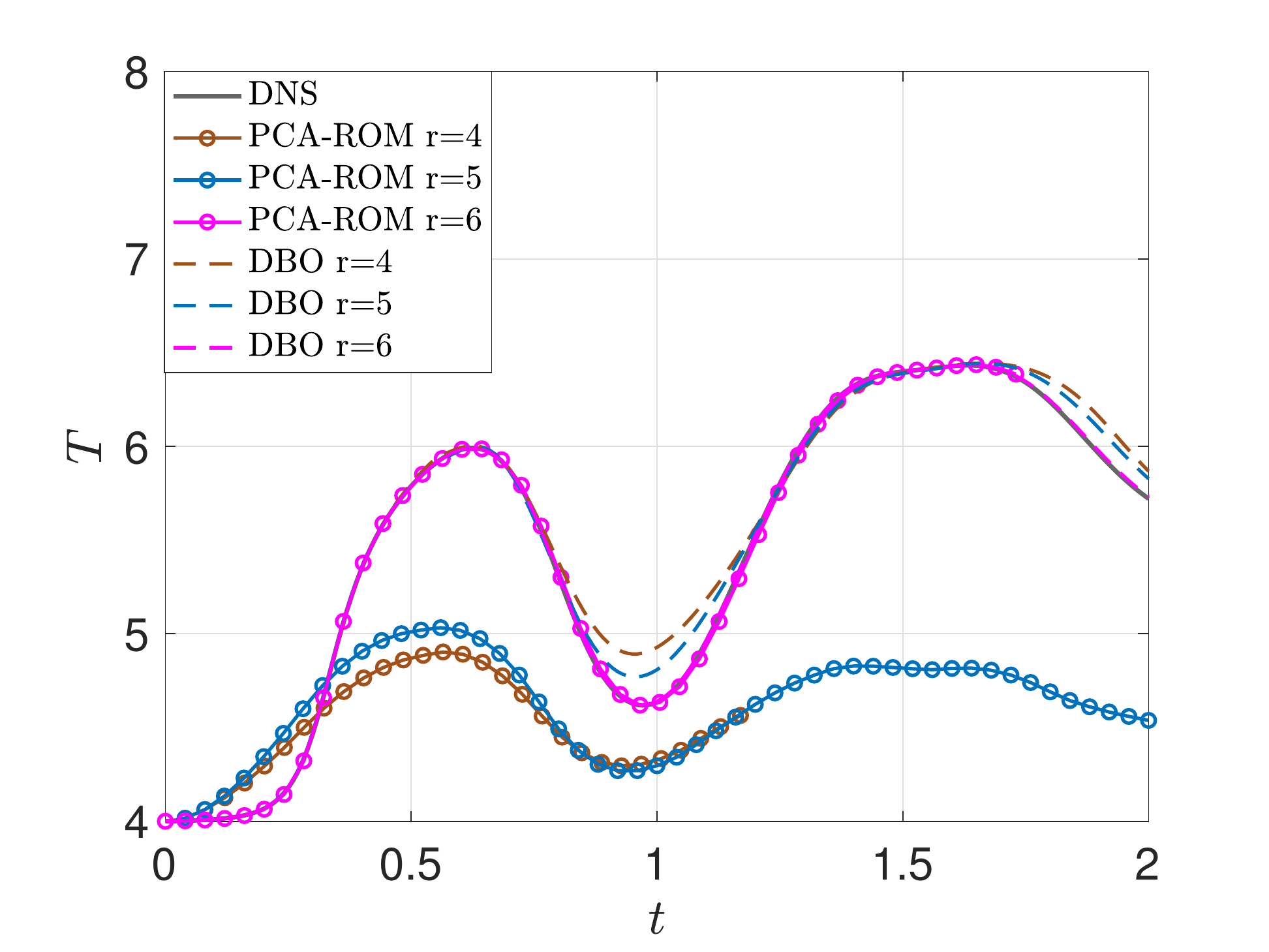}}
 \end{subfigmatrix} \caption{Compressible reactive flow.   Comparing the performance of PCA and DBO using  $r=4,5,$ and $6$ against DNS based on predicted temperature at (a) point 1 and (b) point 2.}
 \label{fig:P1P2-Temp}
\end{figure}

\begin{figure}[!h]
\begin{center}
 \includegraphics[width=\textwidth]{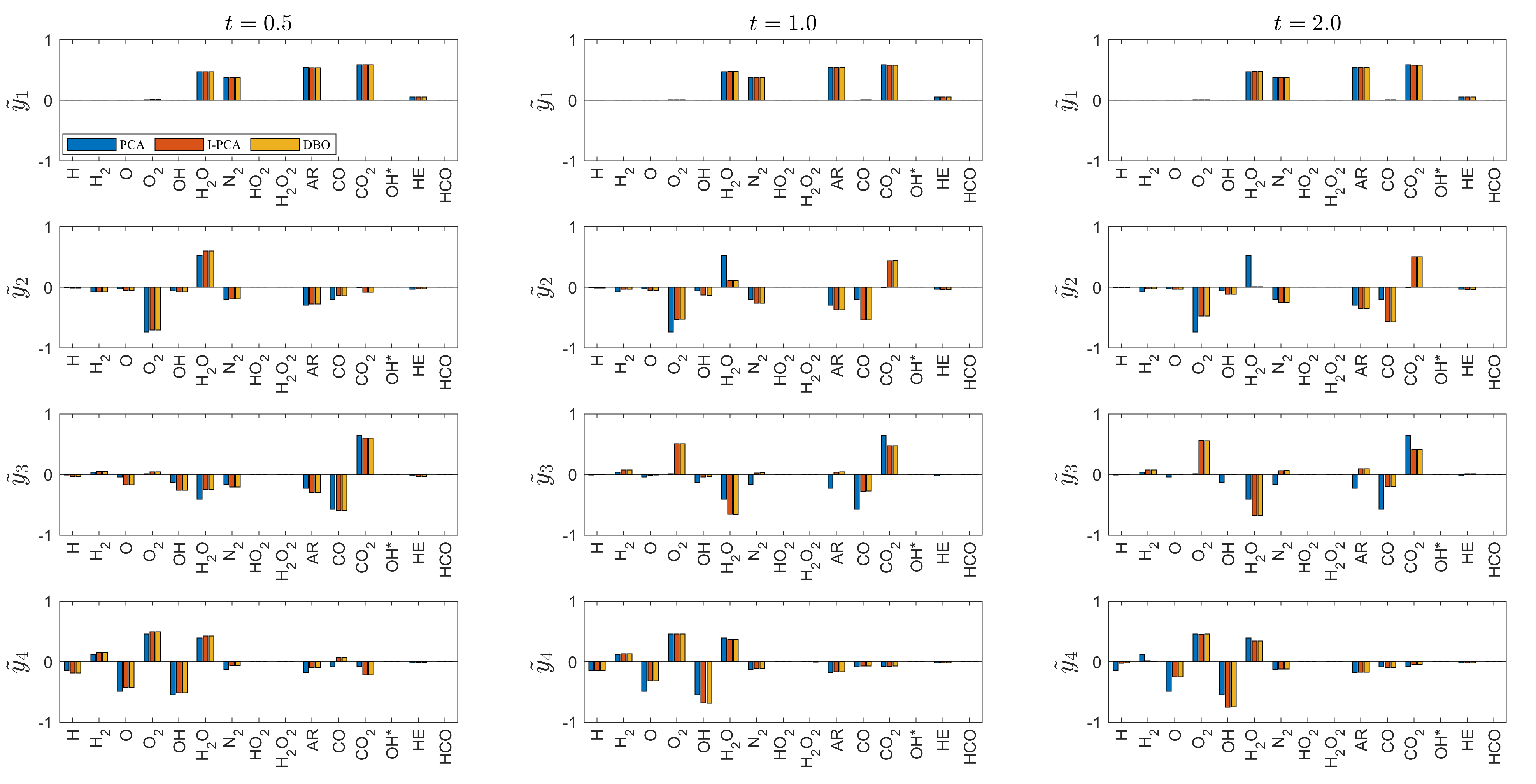}
 \caption{Low-dimensional time-dependent manifolds for compressible reactive flow. Comparison of  species modes obtained from PCA (static), I-PCA and DBO (time-dependent) in different times.  }
 \label{fig:PCA-Temperature}
 \end{center}
\end{figure}

Figure\ \ref{fig:P1P2-Temp} demonstrates the time-plot of temperature at two fixed spatial positions in the domain These two points, labeled as  P1 and P2, are shown  in Figure\ \ref{fig:Comp_NS}(a). P1 is the center of the domain and initially has the fuel composition, and P2 is very close to the shear layer on fuel side. Here, the variations in DNS temperatures after $t=0.5$ are because of passing vortexes. These variations start earlier in P2 with larger amplitudes. PCA-ROM cannot provide a good estimation for temperatures at P1 and P2 with $r<6$, while DBO reasonably estimates the DNS temperature trends  with $r=4$. Figure\ \ref{fig:PCA-Temperature} shows the differences between $\tilde{Y}$ modes in PCA-ROM and DBO as described in \S\ref{subsection:DBO-PCA}. The first species mode of DBO $\tilde{y}_1(t)$ barely changes in time, and it is very similar to the first PCA-ROM species mode $\tilde{y}_1$. This mode is associated with the products of combustion, which occupy a large portion of the domain  and because of this the singular value of the first DBO mode $\hat{\sigma}_1$ is an order of magnitude larger than the next mode, as shown in Fig.\ \ref{fig:Ns-Error}(a). $Y_{DBO}$ and $Y_{PCA}$ have similar patterns in $t=0.5$ but their differences grow significantly in time and as a result PCA-ROM cannot capture the temperature profile in Fig.\ \ref{fig:P1P2-Temp} with $r \le 5$. This means that the proper species composition at each spatial point would not remain in space of the first 5 modes of PCA-ROM during ignition. The contribution of $CO_2$, for example,  in the second and third species modes reveals  the difference between PCA-ROM and DBO. In DBO, $CO_2$ component  in $\tilde{y}_2(t)$ continuously increases from  before ignition ($t=0.5$) to after ignition ($t=2.0$). This can be contrasted against  the static component of$CO_2$ in the  second PCA mode. Similar observations can be made for other species \textit{e.g.} $H_2O$ and $CO$. 

% $\tilde{y}_2(t)$ and $\tilde{y}_3(t)$ have close eigenvalues (Fig.\ \ref{fig:Ns-Error}(a)) and the contribution of similar species are important in these modes, \textit{i.e.} the effects of $CO$ and $CO_2$ are noticeable in $\tilde{y}_2(t)$ and $\tilde{y}_3(t)$.  $OH$, $O$, and $H$ seem to be the important species for the forth mode $\tilde{y}_4(t)$.

\section{Summary} \label{Summary}

In this paper, we presented a variational principle for the determination of a low-rank decomposition (DBO) of the passive and reactive species transport equation.  The optimality conditions of the  variational principle lead to closed-form evolution equations for the components of the decomposition.  The  DBO decomposition consists of  a set of time-dependent orthonormal spatial modes, a set of orthonormal  species modes and a low-rank factorization of the correlation matrix. The novelty of DBO is that all three components are time dependent -- enabling the spatial and species subspaces adapt to the changes of the dynamics on the fly.  The  DBO  decomposition does not require the offline step of generating high-fidelity data  to extract the  low rank composition space. This step is needed in data-driven reduction techniques such as PCA. Instead, the low-rank composition space is extracted from the species transport  equation. Therefore, the DBO decomposition is not  fine-tuned for a target problem and it can extract the low-rank structure for the problem at hand. 

We demonstrated the numerical performance of  DBO  for passive species transport as well as reactive incompressible and compressible flow. In all demonstration cases, we show that the DBO decomposition closely approximates the best instantaneous low-rank decomposition obtained by performing instantaneous PCA of the full-rank species.

\section*{Acknowledgements} \label{Acknowl}
This work has been  supported by the National Science Foundation under Grant No. 2042918 and by NASA under Grant No. 80NSSC18M0150. The authors thank  Prof. Peyman Givi, Prof. Harsha Chelliah and Dr. Jackie Chen for useful comments that led to a number of improvements. The authors also thank  Prerna Patil for her technical input. 

%%%%%%%%%%%%%%%%%%%%%%%%%%%%
%%%-------------------------------------------------------------------------------------------------------------------------
\FloatBarrier
\appendix
\section{Variational Principle for DBO} \label{apnA}
For the sake of simplicity in notation, we denote  $u_i(x,t)$ as $u_i$, $y_i(t)$ as $y_i$ and $\Sigma_{ij}(t)$ as $\Sigma_{ij}$. In the following we show that the first-order optimality condition of the variational principle leads to closed-form evolution equations for $U$, $Y$ and $\Sigma$. Throughout this derivation, we use index notation, where the repeated index implies summation over that index. We begin with the functional form of the variational principle:  
\begin{equation}\label{eq:g}
\begin{aligned}
    \mathcal{G}(\dot{U}, \dot{\Sigma}, \dot{Y},\lambda,\theta) & =  \inner{\dot{u}_{i}}{\dot{u}_{k}}(\Sigma_{ij}\Sigma_{kl})({y_{j}^{T}} {y_{l}}) \\
    & + \inner{u_{i}}{u_{k}} (\dot\Sigma_{ij}\dot\Sigma_{kl})({y_{j}^{T}}{y_{l}}) \\ 
    & + \inner{u_{i}}{u_{k}} (\Sigma_{ij}\Sigma_{kl})(\dot{y}_{j}^{T}\dot{y_{l}})\\
    & +2 \inner{\dot{u}_{i}}{u_{k}}(\Sigma_{ij}\dot\Sigma_{kl})({y_{j}^{T}}{y_{l}})\\
    & +2 \inner{u_{i}}{u_{k}}(\Sigma_{ij}\dot\Sigma_{kl})({y}_{j}^{T}\dot{y_{l}})\\
    & +2 \inner{\dot{u}_{i}}{u_{k}}(\Sigma_{ij}\Sigma_{kl})({y}_{j}^{T}\dot{y_{l}})\\
    & -2\inner{\dot{u}_{i}}{{\mathcal{M}}(\Phi)\Sigma_{ij}{y}_{j}}\\
    & -2\inner{{u}_{i}}{{\mathcal{M}}(\Phi)\dot\Sigma_{ij}{y}_{j}}\\
    & -2\inner{{u}_{i}}{{\mathcal{M}}(\Phi)\Sigma_{ij}\dot{y}_{j}}\\
    & +\big\| {\mathcal{M}}(\Phi)\big\|_{\mathcal{F}}^2 \\
    & + \lambda_{ij}(\inner{{u}_{i}}{\dot{u}_{j}} - \varphi_{ij}) + \gamma_{ij}(y_{i}^{T} \dot{y}_{j} - \theta_{ij}).
  \end{aligned}
\end{equation}
The first-order optimality condition requires a vanishing gradient of the functional $\mathcal{G}$ with respect to all control variables. Starting with the gradient of $\mathcal{G}$ with respect to $\dot{\Sigma}_{mn}$. The following is obtained:
\begin{equation}\label{eq:Sigma-Evo}
\begin{aligned}
      \frac{\partial \mathcal{G}}{\partial \dot{\Sigma} _{mn}} & =  \inner{{u}_{m}}{{u}_{k}}\dot\Sigma_{kl}({y_{n}^{T}} {y_{l}}) + \inner{{u}_{i}}{{u}_{m}}\dot\Sigma_{ij}({y_{j}^{T}} {y_{n}}) \\
     & +  2\inner{\dot{u}_{i}}{{u}_{m}}\Sigma_{ij}({y_{j}^{T}}{{y}_{n}}) \\
     & +  2\inner{{u}_{i}}{{u}_{m}}\Sigma_{ij}({y_{j}^{T}}{\dot{y}_{n}}) \\
     & -2\inner{{u}_{m}}{{\mathcal{M}}(\Phi){y}_{n}} = 0. 
\end{aligned}
\end{equation}
Using the orthonormality conditions and definitions of $\varphi$ and $\theta$, we have:  $\inner{{u}_{m}}{{u}_{k}} = \delta_{mk}$, $ \inner{{u}_{i}}{{u}_{m}}=\delta_{im}$ and ${y_{n}^{T}} {y_{l}} = \delta_{nl}$, ${y_{j}^{T}} {y_{n}}=\delta_{jn}$, $\inner{\dot{u}_{i}}{{u}_{m}}=\varphi_{mi}$ and ${y_{j}^{T}}{\dot{y}_{n}}=\theta_{jn}$ and after simplifications, the evolution equation for  $\Sigma$ is obtained:
\begin{equation}\label{eq:Sigma-2}
\frac{d \Sigma_{mn}}{d t} = \left<u_m , \mathcal{M}(\Phi)y_n\right> - \varphi_{mi}\Sigma_{in}-\Sigma_{mi}\theta_{in}.
\end{equation}
Next, the gradient of $\mathcal{G}$ with respect to $\dot{y}_m$ is set to zero. This results in:
\begin{equation}\label{eq:y-1}
\begin{aligned}
    \frac{\partial \mathcal{G}}{\partial \dot{y} _{m}} & =   2\inner{u_{i}}{{u}_{k}}(\Sigma_{ij}\Sigma_{km})\dot{y}_{j}^{T}\\
    & + 2\inner{{u}_{i}}{{u}_{k}}(\Sigma_{ij}\dot{\Sigma}_{km}){y}_{j}^{T}\\\
    & + 2\inner{\dot{u}_{i}}{{u}_{k}}(\Sigma_{ij}\Sigma_{km}){y_{j}^{T}}\\ 
    & - 2\inner{{u}_{i}}{\Sigma_{im}{\mathcal{M}}(\Phi)}\\
    & + \gamma_{im}y^T_{i} =0.
  \end{aligned}
\end{equation}
To eliminate $\gamma_{im}$, we take the inner product of the above equation and $y_n$ by multiplying $y_n$ from right:
\begin{equation}\label{eq:y-2}
\begin{aligned}
\frac{\partial \mathcal{G}}{\partial \dot{y} _{m}} y_{n}& = 2\delta_{ik}(\Sigma_{ij}\Sigma_{km})\theta_{nj}\\
    & + 2\delta_{ik}(\Sigma_{ij}\dot{\Sigma}_{km})\delta_{jn}\\\
    & + 2\varphi_{ki}(\Sigma_{ij}\Sigma_{km})\delta_{jn}\\ 
    & - 2\inner{{u}_{i}}{\Sigma_{im}{\mathcal{M}}(\Phi)y_n}\\
    & + \gamma_{im}\delta_{in} =0.
% \frac{\partial \mathcal{G}}{\partial \dot{y} _{m}}   y_{n}& =      2\inner{\dot{u}_{i}}{{u}_{k}}(\Sigma_{ij}\Sigma_{km})\delta_{jn} \\
%     & + 2\inner{{u}_{i}}{{u}_{k}}(\dot{\Sigma}_{ij}\Sigma_{km})\delta_{jn} \\ 
%     & - 2\inner{u_{i}}{\Sigma_{im}{\mathcal{M}}(\Phi)y_{n}}\\
%     & + \gamma_{jm}\delta_{jn} =0.
  \end{aligned}
\end{equation}
 Then $\gamma_{nm}$ is obtained from: 
\begin{equation}\label{eq:gamma}
\begin{aligned}
\gamma_{nm} & =  2\inner{u_{i}}{\Sigma_{im}{\mathcal{M}}(\Phi)y_{n}}   \\
            & -2 \Sigma_{ij}\Sigma_{im}\theta_{nj}\\
            & - 2\Sigma_{in}\dot{\Sigma}_{im} \\ 
            & - 2\varphi_{ki}\Sigma_{in}\Sigma_{km}. \\
  \end{aligned}
\end{equation}
By Substituting $\gamma_{nm}$ from Eq.\ (\ref{eq:gamma}) into Eq.\ (\ref{eq:y-1}) and simplifying, We obtain:
\begin{equation}\label{eq:y-3}
\begin{aligned}
\Sigma_{ij}\dot{y}^T_j  =  \inner{u_{i}}{\mathcal{M}(\Phi)}(I-y_n y^T_n) + \Sigma_{ij}\theta_{nj}y^T_n,
  \end{aligned}
\end{equation}
where $I \in \mathbb{R}^{n_s \times n_s}$ is the identity matrix. Taking the transpose of the Eq.\ (\ref{eq:y-3}) results in:
\begin{equation}\label{eq:y-4}
\begin{aligned}
\Sigma_{ji}\dot{y}_j  = (I-y_n y^T_n) \inner{\mathcal{M}(\Phi)}{u_{i}} + \Sigma_{ji}\theta_{nj}y_n,
  \end{aligned}
\end{equation}
where we have used: $\inner{u_{i}}{\mathcal{M}(\Phi)}^T = \inner{\mathcal{M}(\Phi)}{u_{i}}$ and $\theta_{nj}^T=-\theta_{jn}$. Dividing both sides of Eq.\ (\ref{eq:y-4}) by the inverse of $\Sigma^T$ yields the evolution equation for the  species modes as in the following:
\begin{equation}\label{eq:ydot}
\begin{aligned}
\dot{y}_j  = (I-y_n y^T_n) \inner{\mathcal{M}(\Phi)}{u_{i}}\Sigma^{-1}_{ji} + y_n\theta_{nj}.
  \end{aligned}
\end{equation}
Next, the gradient of $\mathcal{G}$ with respect to $\dot{u}_m$ is considered. Since $u$ is a  function, \textit{i.e.} infinite-dimensional, we use  the Fr\'{e}chet differential:
\begin{equation*}
\mathcal{G}'|_{\dot{U}} \triangleq \lim_{\epsilon \rightarrow 0 } \frac{\mathcal{G}(\dot{U}+\epsilon \dot{U}',\dot{\Sigma},\dot{Y},\lambda,\gamma)-\mathcal{G}(\dot{U},\dot{\Sigma},\dot{Y},\lambda,\gamma)}{\epsilon}.
\end{equation*}
For optimality condition the Fr\'{e}chet differential of $\mathcal{G}$ with respect to every column of $\dot{U}$ must vanish. This results in:
\begin{equation}\label{eq:Frechet-u}
\begin{aligned}
   \mathcal{G}'|_{\dot{u}_m}  & =  2 \inner{\dot{u}'}{\dot{u}_{k}} (\Sigma_{mj}\Sigma_{kl})({y_{j}^{T}}{y_{l}}) \\
    & + 2 \inner{\dot{u}'}{{u_{k}}} (\Sigma_{mj}\dot\Sigma_{kl})({y_{j}^{T}}{y_{l}}) \\ 
    & + 2 \inner{\dot{u}'}{ {u_{k}}}(\Sigma_{mj}\Sigma_{kl})({y}_{j}^{T}\dot{y_{l}})\\
    & - 2 \inner{\dot{u}'}{\Sigma_{mj}{\mathcal{M}}(\Phi)y_{j}}\\
    & + \lambda_{im} \inner{\dot{u}'}{u_{i}} =0.
  \end{aligned}
\end{equation}
The above equation can be written as $\inner{\dot{u}'}{\nabla_{\dot{u}_m} \mathcal{G}}$ and since   $\dot{u}'$ is an arbitrary perturbation, $\nabla_{\dot{u}_m} \mathcal{G}$ must vanish. This  results in:
\begin{equation}\label{eq:u-1}
\begin{aligned}
    \nabla_{\dot{u}_m} \mathcal{G} & =  2\dot{u}_{k}(\Sigma_{mj}\Sigma_{kl})({y_{j}^{T}}{y_{l}}) \\
    & + 2{u_{k}}(\Sigma_{mj}\dot\Sigma_{kl})({y_{j}^{T}}{y_{l}}) \\ 
    & + 2{u_{k}}(\Sigma_{mj}\Sigma_{kl})({y}_{j}^{T}\dot{y_{l}})\\
    & - 2\Sigma_{mj}{\mathcal{M}}(\Phi)y_{j}\\
    & + \lambda_{im}u_{i} =0.
  \end{aligned}
\end{equation}
Similar to the procedure of deriving the evolution equation for $\dot{y}_m$, to eliminate  $\lambda_{im}$ an inner product of Eq.\ (\ref{eq:u-1}) and $u_n$  is taken. This results in:
\begin{equation}\label{eq:u-2}
\begin{aligned}
 \inner{{u}_{n}}{\nabla_{\dot{u}_m} \mathcal{G}} &= 
 2\varphi_{nk}\Sigma_{mj}\Sigma_{kl}\delta_{jl}\\
 & +2\delta_{nk}(\Sigma_{mj}\dot\Sigma_{kl})\delta_{jl} \\
    & + 2\delta_{nk}(\Sigma_{mj}\Sigma_{kl})\theta_{jl}\\ 
    & - 2\inner{u_{n}}{\Sigma_{mj}{\mathcal{M}}(\Phi)y_{j}}\\
    & + \lambda_{im} \delta_{in}=0.
  \end{aligned}
\end{equation}
Solving for $\lambda_{nm}$ using Eq.\ (\ref{eq:u-2})  results in: 
\begin{equation}\label{eq:lambda}
\begin{aligned}
 \lambda_{nm}  &=   2\inner{u_{n}}{\Sigma_{mj}{\mathcal{M}}(\Phi)y_{j}}\\
 & -2 \varphi_{nk} \Sigma_{mj} \Sigma_{kj}\\
  &-2\Sigma_{mj}\dot\Sigma_{nj} \\
    & -2\Sigma_{mj}\Sigma_{nl} \theta_{jl}.\\
  \end{aligned}
\end{equation}
Replacing $\lambda_{mn}$ from Eq.\ (\ref{eq:lambda}) into Eq.\ (\ref{eq:u-1}) and simplifying the results yields:
\begin{equation}\label{eq:u-3}
\begin{aligned}
\dot{u}_i \Sigma_{ij} =  (\mathcal{M}(\Phi)y_{j} - u_n\inner{u_n}{\mathcal{M}(\Phi)y_{j}}) + u_n\varphi_{ni}\Sigma_{ij}.
  \end{aligned}
\end{equation}
Multiplying both sides of Eq.\ (\ref{eq:u-3}) by the inverse of $\Sigma$ from right results in:
\begin{equation}\label{eq:udot}
\begin{aligned}
 \dot{u}_i   =  (\mathcal{M}(\Phi)y_{j} - u_n\inner{u_n}{\mathcal{M}(\Phi)y_{j}})\Sigma^{-1}_{ij} + u_n\varphi_{ni},
  \end{aligned}
\end{equation}
where $\Sigma^{-1}_{ij}$ is the $ij$ element of matrix $\Sigma^{-1}$. Therefore, evolution equations of DBO components are as in the following:
\begin{align}
%  \frac{\partial {u}_i}{\partial t}  &=   (\mathcal{M}(\Phi)y_{j} - u_n\inner{u_n}{\mathcal{M}(\Phi)y_{j}})\Sigma^{-1}_{ji} + u_n\varphi_{ni}, \label{eq:dudt} \\
 \frac{\partial U}{\partial t}  &=   (\mathcal{M}(\Phi)Y - U\inner{U}{\mathcal{M}(\Phi)Y})\Sigma^{-1} + U\varphi, \label{eq:dudt_gen} \\
 \frac{d \Sigma}{d t} &= \left<U , \mathcal{M}(\Phi)Y\right>  - \varphi \Sigma-\Sigma \theta, \label{eq:dSdt_gen} \\
 \frac{dY}{dt}  &= (I-Y Y^T) \inner{\mathcal{M}(\Phi)}{U}\Sigma^{-T} + Y\theta. \label{eq:dydt_gen}
\end{align}

\section{Equivalence of DBO Decomposition} \label{apnB}
Any choice of skew-symmetric matrices for $\varphi$ and $\theta$ lead to equivalent decompositions. Two DBO decompositions are equivalent if they represent the same low-rank subspace instantaneously. Therefore, the two DBO decompositions $\{U,\Sigma,Y\}$ and $\{\hat{U},\hat{\Sigma},\hat{Y}\}$ are equivalent if and only if: $U\Sigma Y^T= \hat{U}\hat{\Sigma}\hat{Y}^T$. As a result, if 
\begin{equation}\label{eq:equiv}
    \hat{U}=U R_U, \quad  \quad\hat{Y}=YR_Y \quad \mbox{and} \quad \hat{\Sigma}= R_U^T\Sigma R_Y,
\end{equation}
  for any orthonormal matrices $R_U \in \mathbb{R}^{r\times r}$ and  $R_Y \in \mathbb{R}^{r\times r}$, then it is straightforward to show that the two decompositions are equivalent using $R_U^TR_U=I$ and $R_YR_Y^T=I$, where $I \in \mathbb{R}^{r\times r}$ is the identity matrix. The matrices $R_U$ and $R_Y$ are \emph{in-subspace rotations}. \textit{i.e.}, both $U$ and $\hat{U}$ span the same subspace. The same is true for $Y$. 

Now, let $\{U,\Sigma,Y\}$ and $\{\hat{U},\hat{\Sigma},\hat{Y}\}$ be the solution of    Eqs.~(\ref{eq:dudtgen}-\ref{eq:dydtgen})  with the skew-symmetric matrices $\{\varphi, \theta \}$ and $\{\hat{\varphi}, \hat{\theta} \}$, respectively. It can be shown that  the two decompositions are equivalent for all $t>0$ given that: (i) they are equivalent at $t=0$ with the corresponding rotation matrices $R_{U_0}$ and $R_{Y_0}$, and (ii) the rotation matrices $R_U$ and $R_Y$ evolve according to:
\begin{equation}\label{eq:Rdot}
    \dot{R}_U = R_U \hat{\varphi} - \varphi R_U \quad \mbox{and} \quad \dot{R}_Y = R_Y \hat{\theta} - \theta R_Y,
\end{equation} 
with the initial conditions of $R_{U_0}$ and $R_{Y_0}$, respectively.   

It was shown in Lemma 2.1 in Ref.~\cite{Babaee_PRSA} that matrices $R_U$ and $R_Y$ remain orthonormal for all $t>0$ if they evolve according to Eqs.~(\ref{eq:Rdot}). To prove the equivalence under the above two conditions, we show that the  DBO decompositions  $\{\hat{U},\hat{\Sigma},\hat{Y}\}$ is an in-subspace rotation of $\{U,\Sigma,Y\}$ according to Eq.~(\ref{eq:equiv}) and the rotation matrices are governed by  Eq.~(\ref{eq:Rdot}). To this end,  we start from the evolution equations for  $\{\hat{U},\hat{\Sigma},\hat{Y}\}$ and using the rotation matrices given by  Eq.~(\ref{eq:Rdot}), we recover the DBO evolution equations for $\{U,\Sigma,Y\}$. The DBO evolution equations for  $\{\hat{U},\hat{\Sigma},\hat{Y}\}$ are given by:
\begin{equation*}
    \dot{\hat{U}} = \underset{\perp \hat{U}}{\prod } \mathcal{M}(\Phi)\hat{Y}\hat{\Sigma}^{-1} + \hat{U}\hat{\varphi}.
\end{equation*}
First we note that: $\underset{\perp \hat{U}}{\prod } = \underset{\perp U}{\prod }$, since 
\begin{equation*}
\underset{\perp \hat{U}}{\prod f} = f-\hat{U}\inner{\hat{U}}{f} =  f-UR_U\inner{UR_U}{f} = f-UR_UR_U^T\inner{U}{f} = f-U\inner{U}{f},
\end{equation*}
for any $f \in \mathbb{R}^{\infty \times 1}$.
Therefore, 
\begin{align*}
    \dot{U}R_U + U\dot{R}_U &= \underset{\perp U}{\prod } \mathcal{M}(\Phi)YR_Y R_Y^{-1} \Sigma^{-1}R_U^{-T} + UR_U\hat{\varphi}= \underset{\perp U}{\prod } \mathcal{M}(\Phi)Y \Sigma^{-1}R_U^{-T} + UR_U\hat{\varphi}.
\end{align*}
Rearranging the above equation and multiplying it by $R_U^T$ from right results in:
\begin{equation*}
    \dot{U} 
    = \underset{\perp U}{\prod } \mathcal{M}(\Phi)Y \Sigma^{-1} + UR_U\hat{\varphi}R_U^T-U\dot{R}_U R_U^T,
\end{equation*}
Now replacing $\dot{R}_U = R_U \hat{\varphi} - \varphi R_U$ in the above equation results in:
\begin{align*}
    \dot{U} 
    &= \underset{\perp U}{\prod } \mathcal{M}(\Phi)Y \Sigma^{-1} + UR_U\hat{\varphi}R_U^T-U(R_U \hat{\varphi} - \varphi R_U)R_U^T\\
    &= \underset{\perp U}{\prod } \mathcal{M}(\Phi)Y \Sigma^{-1} + U \varphi,
\end{align*}
which is the governing equation for $U$. The analogous procedure
can be repeated for the evolution equation of $Y$.  Replacing the equivalence relationship for the evolution of $\hat{\Sigma}$ results in:

\begin{align*}
    \dot{R}_U^T\Sigma R_Y +R_U^T\dot{\Sigma}R_Y+R_U^T\Sigma\dot{R}_Y &=\inner{UR_U}{\mathcal{M}(\Phi) Y R_Y} - \hat{\varphi}R_U^T\Sigma R_Y - R_U^T\Sigma R_Y\hat{\theta}\\
    & = R_U^T\inner{U}{\mathcal{M}(\Phi) Y }R_Y - \hat{\varphi}R_U^T\Sigma R_Y - R_U^T\Sigma R_Y\hat{\theta}.
\end{align*}
Multiplying the above equation from left by $R_U$ and from right by $R_Y^T$ and rearranging to obtain the evolution equation for $\dot{\Sigma}$ results in:
\begin{align*}
    \dot{\Sigma}=\inner{U}{\mathcal{M}(\Phi)} - R_U\hat{\varphi}R_U^T\Sigma - \Sigma R_Y\hat{\theta}R_Y^T - R_U\dot{R}_U^T\Sigma -\Sigma\dot{R}_YR_Y^T.
\end{align*}
The term $\dot{R}_U^T$ can be obtained from  Eq. (\ref{eq:Rdot}):  $\dot{R}_U^T=\varphi^TR_U^T-R_U^T\varphi^T=-\varphi R_U^T+R_U^T\varphi$, where we have used $\varphi^T=-\varphi$. Using this relation for  $\dot{R}_U^T$ and replacing $\dot{R}_Y$ from  Eq. (\ref{eq:Rdot})  into the above equation and after simplification yields the evolution equation for $\dot{\Sigma}$, \textit{i.e.}, Eq. (\ref{eq:dSdtgen}).

\bibliography{mybibfile,Hessam}

\begin{thebibliography}{10}
\expandafter\ifx\csname url\endcsname\relax
  \def\url#1{\texttt{#1}}\fi
\expandafter\ifx\csname urlprefix\endcsname\relax\def\urlprefix{URL }\fi
\expandafter\ifx\csname href\endcsname\relax
  \def\href#1#2{#2} \def\path#1{#1}\fi

\bibitem{LL09}
T.~Lu, C.~Law, {T}oward {A}ccommodating {R}ealistic {F}uel {C}hemistry in
  {L}arge-{S}cale {C}omputations, Prog. Energy Combust. Sci. 35~(2) (2009)
  192--215.

\bibitem{johnson1995}
G.~T. Johnson, L.~J. Hunter, A numerical study of dispersion of passive scalars
  in city canyons, Boundary-Layer Meteorology 75~(3) (1995) 235--262.

\bibitem{warhaft2000}
Z.~Warhaft, Passive scalars in turbulent flows, Annual Review of Fluid
  Mechanics 32~(1) (2000) 203--240.

\bibitem{anand2003}
M.~Anand, K.~Rajagopal, K.~Rajagopal, A model incorporating some of the
  mechanical and biochemical factors underlying clot formation and dissolution
  in flowing blood, Journal of Theoretical Medicine 5~(3-4) (2003) 183--218.

\bibitem{antonia2003}
R.~Antonia, P.~Orlandi, Effect of schmidt number on small-scale passive scalar
  turbulence, Appl. Mech. Rev. 56~(6) (2003) 615--632.

\bibitem{SM06}
C.~Safta, C.~Madnia, {A}utoignition and {S}tructure of {N}onpremixed {CH4/H2}
  {F}lames: {D}etailed and {R}educed {K}inetic {M}odels, Combust. Flame
  144~(1-2) (2006) 64--73.

\bibitem{li2009}
X.-X. Li, C.-H. Liu, D.~Y. Leung, Numerical investigation of pollutant
  transport characteristics inside deep urban street canyons, Atmospheric
  Environment 43~(15) (2009) 2410--2418.

\bibitem{wang2015}
H.~Wang, M.~Sun, Y.~Yang, N.~Qin, A passive scalar-based method for numerical
  combustion, International Journal of Hydrogen Energy 40~(33) (2015)
  10658--10661.

\bibitem{cerrolaza2017}
M.~Cerrolaza, S.~Shefelbine, D.~Garz{\'o}n-Alvarado, Numerical methods and
  advanced simulation in biomechanics and biological processes, Academic Press,
  2017.

\bibitem{P13}
S.~B. Pope,
  \href{http://www.sciencedirect.com/science/article/pii/S1540748912003963}{Small
  scales, many species and the manifold challenges of turbulent combustion},
  Proceedings of the Combustion Institute 34~(1) (2013) 1--31.
\newblock \href {http://dx.doi.org/https://doi.org/10.1016/j.proci.2012.09.009}
  {\path{doi:https://doi.org/10.1016/j.proci.2012.09.009}}.
\newline\urlprefix\url{http://www.sciencedirect.com/science/article/pii/S1540748912003963}

\bibitem{SYWL09}
D.~A. Sheen, X.~You, H.~Wang, T.~L{\o}v{\aa}s, {S}pectral {U}ncertainty
  {Q}uantification, {P}ropagation and {O}ptimization of a {D}etailed {K}inetic
  {M}odel for {E}thylene {C}ombustion, Proc. Combust. Inst. 32~(1) (2009)
  535--542.

\bibitem{L07}
C.~K. Law,
  \href{http://www.sciencedirect.com/science/article/pii/S1540748906003877}{Combustion
  at a crossroads: Status and prospects}, Proceedings of the Combustion
  Institute 31~(1) (2007) 1--29.
\newblock \href {http://dx.doi.org/https://doi.org/10.1016/j.proci.2006.08.124}
  {\path{doi:https://doi.org/10.1016/j.proci.2006.08.124}}.
\newline\urlprefix\url{http://www.sciencedirect.com/science/article/pii/S1540748906003877}

\bibitem{coltrin2003}
M.~E. Coltrin, P.~Glarborg, Chemically reacting flow: theory and practice,
  Wiley-Interscience, 2003.

\bibitem{NGL19}
A.~Nouri, P.~Givi, D.~Livescu, {M}odeling and {S}imulation of {T}urbulent
  {N}uclear {F}lames in {T}ype {I}a {S}upernovae, Prog. Aerosp. Sci. 108 (2019)
  156--179.

\bibitem{CFD2030}
J.~Slotnick, A.~Khodadoust, J.~Alonso, D.~Darmofal, W.~Gropp, E.~Lurie,
  D.~Mavriplis, {CFD} {V}ision 2030 {S}tudy: {A} {P}ath to {R}evolutionary
  {C}omputational {A}erosciences.

\bibitem{BAB12}
J.~C. Bennett, H.~Abbasi, P.~Bremer, R.~Grout, A.~Gyulassy, T.~Jin, S.~Klasky,
  H.~Kolla, M.~Parashar, V.~Pascucci, P.~Pebay, D.~Thompson, H.~Yu, F.~Zhang,
  J.~Chen, Combining in-situ and in-transit processing to enable extreme-scale
  scientific analysis, in: SC '12: Proceedings of the International Conference
  on High Performance Computing, Networking, Storage and Analysis, 2012, pp.
  1--9.
\newblock \href {http://dx.doi.org/10.1109/SC.2012.31}
  {\path{doi:10.1109/SC.2012.31}}.

\bibitem{AM_DOE_14}
Applied mathematics research for exascale computing, U.S. Department of Energy,
  Office of Science, Advanced Scientific Computing Research Program.

\bibitem{VT86}
S.~Vajda, T.~Tur{\'a}nyi, {P}rincipal {C}omponent {A}nalysis for {R}educing the
  {E}delson-{F}ield-{N}oyes {M}odel of the {B}elousov-{Z}habotinskii
  {R}eaction, J. Phys. Chem. 90~(8) (1986) 1664--1670.

\bibitem{EC11}
G.~Esposito, H.~Chelliah, {S}keletal {R}eaction {M}odels based on {P}rincipal
  {C}omponent {A}nalysis: {A}pplication to {E}thylene--{A}ir {I}gnition,
  {P}ropagation, and {E}xtinction {P}henomena, Combust. Flame 158~(3) (2011)
  477--489.

\bibitem{SFCFR16}
A.~Stagni, A.~Frassoldati, A.~Cuoci, T.~Faravelli, E.~Ranzi, {S}keletal
  {M}echanism {R}eduction {T}hrough {S}pecies-{T}argeted {S}ensitivity
  {A}nalysis, Combust. Flame 163 (2016) 382--393.

\bibitem{lu2005}
T.~Lu, C.~K. Law, A directed relation graph method for mechanism reduction,
  Proceedings of the Combustion Institute 30~(1) (2005) 1333--1341.

\bibitem{pepiot2008}
P.~Pepiot-Desjardins, H.~Pitsch, An efficient error-propagation-based reduction
  method for large chemical kinetic mechanisms, Combust. Flame 154~(1-2) (2008)
  67--81.

\bibitem{niemeyer2010}
K.~E. Niemeyer, C.-J. Sung, M.~P. Raju, Skeletal mechanism generation for
  surrogate fuels using directed relation graph with error propagation and
  sensitivity analysis, Combustion and flame 157~(9) (2010) 1760--1770.

\bibitem{Elliott04}
L.~Elliott, D.~Ingham, A.~Kyne, N.~Mera, M.~Pourkashanian, C.~Wilson, {G}enetic
  {A}lgorithms for {O}ptimisation of {C}hemical {K}inetics {R}eaction
  {M}echanisms, Prog. Energy Combust. Sci. 30~(3) (2004) 297--328.

\bibitem{Sikalo14}
N.~Sikalo, O.~Hasemann, C.~Schulz, A.~Kempf, I.~Wlokas, {A} {G}enetic
  {A}lgorithm-{B}ased {M}ethod for the {A}utomatic {R}eduction of {R}eaction
  {M}echanisms, Int. J. Chem. Kinet. 46~(1) (2014) 41--59.

\bibitem{Djouad03}
R.~Djouad, B.~Sportisse, N.~Audiffren, {R}eduction of {M}ultiphase
  {A}tmospheric {C}hemistry, J. Atmos. Chem. 46~(2) (2003) 131--157.

\bibitem{Gao16}
Y.~Gao, R.~Shan, S.~Lyra, C.~Li, H.~Wang, J.~Chen, T.~Lu, {O}n
  {L}umped-{R}educed {R}eaction {M}odel for {C}ombustion of {L}iquid {F}uels,
  Combust. Flame 163 (2016) 437--446.

\bibitem{Stiefenhofer98}
M.~Stiefenhofer, {Q}uasi-{S}teady-{S}tate {A}pproximation for {C}hemical
  {R}eaction {N}etworks, J. Math. Biol. 36~(6) (1998) 593--609.

\bibitem{Rein92}
M.~Rein, {T}he {P}artial-{E}quilibrium {A}pproximation in {R}eacting {F}lows,
  Phys. Fluids A: Fluid Dyn. 4~(5) (1992) 873--886.

\bibitem{Keck90}
J.~Keck, {R}ate-{C}ontrolled {C}onstrained-{E}quilibrium {T}heory of {C}hemical
  {R}eactions in {C}omplex {S}ystems, Prog. Energy Combust. Sci. 16~(2) (1990)
  125--154.

\bibitem{LG89}
S.~Lam, D.~Goussis, {U}nderstanding {C}omplex {C}hemical {K}inetics with
  {C}omputational {S}ingular {P}erturbation, in: Symp. Combust. Proc., Vol.~22,
  Elsevier, 1989, pp. 931--941.

\bibitem{GIV13}
S.~Gupta, H.~Im, M.~Valorani, {A}nalysis of n-{H}eptane {A}uto-{I}gnition
  {C}haracteristics using {C}omputational {S}ingular {P}erturbation, Proc.
  Combust. Inst. 34~(1) (2013) 1125--1133.

\bibitem{MP92}
U.~Maas, S.~Pope, {S}implifying {C}hemical {K}inetics: {I}ntrinsic
  {L}ow-{D}imensional {M}anifolds in {C}omposition {S}pace, Combust. Flame
  88~(3-4) (1992) 239--264.

\bibitem{SP09}
J.~C. Sutherland, A.~Parente,
  \href{http://www.sciencedirect.com/science/article/pii/S1540748908002630}{Combustion
  modeling using principal component analysis}, Proceedings of the Combustion
  Institute 32~(1) (2009) 1563--1570.
\newblock \href {http://dx.doi.org/https://doi.org/10.1016/j.proci.2008.06.147}
  {\path{doi:https://doi.org/10.1016/j.proci.2008.06.147}}.
\newline\urlprefix\url{http://www.sciencedirect.com/science/article/pii/S1540748908002630}

\bibitem{ME15}
T.~Mirgolbabaei, H.and~Echekki,
  \href{http://www.sciencedirect.com/science/article/pii/S0010218014003812}{The
  reconstruction of thermo-chemical scalars in combustion from a reduced set of
  their principal components}, Combustion and Flame 162~(5) (2015) 1650--1652.
\newblock \href
  {http://dx.doi.org/https://doi.org/10.1016/j.combustflame.2014.11.027}
  {\path{doi:https://doi.org/10.1016/j.combustflame.2014.11.027}}.
\newline\urlprefix\url{http://www.sciencedirect.com/science/article/pii/S0010218014003812}

\bibitem{OE17}
O.~Owoyele, T.~Echekki, {T}oward {C}omputationally {E}fficient {C}ombustion
  {DNS} with {C}omplex {F}uels via {P}rincipal {C}omponent {T}ransport,
  Combust. Theory Model 21~(4) (2017) 770--798.

\bibitem{MOCP20}
M.~R. Malik, P.~Obando~Vega, A.~Coussement, A.~Parente,
  \href{http://www.sciencedirect.com/science/article/pii/S1540748920301966}{Combustion
  modeling using principal component analysis: A posteriori validation on
  sandia flames d, e and f}, Proceedings of the Combustion Institute\href
  {http://dx.doi.org/https://doi.org/10.1016/j.proci.2020.07.014}
  {\path{doi:https://doi.org/10.1016/j.proci.2020.07.014}}.
\newline\urlprefix\url{http://www.sciencedirect.com/science/article/pii/S1540748920301966}

\bibitem{SL09}
T.~Sapsis, P.~Lermusiaux, Dynamically orthogonal field equations for continuous
  stochastic dynamical systems, Physica D: Nonlinear Phenomena 238~(23-24)
  (2009) 2347--2360.

\bibitem{CHZI13}
M.~Cheng, T.~Y. Hou, Z.~Zhang,
  \href{http://www.sciencedirect.com/science/article/pii/S0021999113001526}{A
  dynamically bi-orthogonal method for time-dependent stochastic partial
  differential equations i: Derivation and algorithms}, Journal of
  Computational Physics 242~(0) (2013) 843 -- 868.
\newblock \href {http://dx.doi.org/http://dx.doi.org/10.1016/j.jcp.2013.02.033}
  {\path{doi:http://dx.doi.org/10.1016/j.jcp.2013.02.033}}.
\newline\urlprefix\url{http://www.sciencedirect.com/science/article/pii/S0021999113001526}

\bibitem{PB20}
P.~Patil, H.~Babaee,
  \href{http://www.sciencedirect.com/science/article/pii/S0021999120302850}{Real-time
  reduced-order modeling of stochastic partial differential equations via
  time-dependent subspaces}, Journal of Computational Physics 415 (2020)
  109511.
\newblock \href {http://dx.doi.org/https://doi.org/10.1016/j.jcp.2020.109511}
  {\path{doi:https://doi.org/10.1016/j.jcp.2020.109511}}.
\newline\urlprefix\url{http://www.sciencedirect.com/science/article/pii/S0021999120302850}

\bibitem{Babaee_PRSA}
H.~Babaee, T.~P. Sapsis, \href{http://dx.doi.org/10.1098/rspa.2015.0779}{A
  minimization principle for the description of modes associated with
  finite-time instabilities}, Proceedings of the Royal Society of London A:
  Mathematical, Physical and Engineering Sciences 472~(2186).
\newline\urlprefix\url{http://dx.doi.org/10.1098/rspa.2015.0779}

\bibitem{CSK14}
M.~Choi, T.~P. Sapsis, G.~E. Karniadakis,
  \href{http://www.sciencedirect.com/science/article/pii/S002199911400237X}{On
  the equivalence of dynamically orthogonal and bi-orthogonal methods: Theory
  and numerical simulations}, Journal of Computational Physics 270 (2014) 1 --
  20.
\newblock \href {http://dx.doi.org/http://dx.doi.org/10.1016/j.jcp.2014.03.050}
  {\path{doi:http://dx.doi.org/10.1016/j.jcp.2014.03.050}}.
\newline\urlprefix\url{http://www.sciencedirect.com/science/article/pii/S002199911400237X}

\bibitem{KL07}
O.~Koch, C.~Lubich, \href{http://dx.doi.org/10.1137/050639703}{Dynamical
  low‐rank approximation}, SIAM Journal on Matrix Analysis and Applications
  29~(2) (2007) 434--454.
\newblock \href {http://dx.doi.org/10.1137/050639703}
  {\path{doi:10.1137/050639703}}.
\newline\urlprefix\url{http://dx.doi.org/10.1137/050639703}

\bibitem{MNZ15}
E.~Musharbash, F.~Nobile, T.~Zhou,
  \href{https://doi.org/10.1137/140967787}{Error analysis of the dynamically
  orthogonal approximation of time dependent random pdes}, SIAM Journal on
  Scientific Computing 37~(2) (2015) A776--A810.
\newblock \href {http://dx.doi.org/10.1137/140967787}
  {\path{doi:10.1137/140967787}}.
\newline\urlprefix\url{https://doi.org/10.1137/140967787}

\bibitem{BFHS17}
H.~Babaee, M.~Farazmand, G.~Haller, T.~P. Sapsis,
  \href{http://dx.doi.org/10.1063/1.4984627}{Reduced-order description of
  transient instabilities and computation of finite-time {L}yapunov exponents},
  Chaos: An Interdisciplinary Journal of Nonlinear Science 27~(6) (2017)
  063103.
\newblock \href {http://dx.doi.org/10.1063/1.4984627}
  {\path{doi:10.1063/1.4984627}}.
\newline\urlprefix\url{http://dx.doi.org/10.1063/1.4984627}

\bibitem{bt2004}
Z.~Battles, L.~N. Trefethen, An extension of matlab to continuous functions and
  operators, SIAM Journal on Scientific Computing 25~(5) (2004) 1743--1770.

\bibitem{DCB20}
M.~Donello, M.~Carpenter, H.~Babaee, Computing sensitivities in evolutionary
  systems: A real-time reduced order modeling strategy (2020).
\newblock \href {http://arxiv.org/abs/2012.14028} {\path{arXiv:2012.14028}}.

\bibitem{B19}
H.~Babaee, \href{https://doi.org/10.1098/rspa.2019.0506}{An observation-driven
  time-dependent basis for a reduced description of transient stochastic
  systems}, Proceedings of the Royal Society A: Mathematical, Physical and
  Engineering Sciences 475~(2231) (2019) 20190506.
\newblock \href {http://dx.doi.org/10.1098/rspa.2019.0506}
  {\path{doi:10.1098/rspa.2019.0506}}.
\newline\urlprefix\url{https://doi.org/10.1098/rspa.2019.0506}

\bibitem{A91}
N.~Aubry,
  \href{https://www.scopus.com/inward/record.uri?eid=2-s2.0-0008458408&doi=10.1007%2fBF00271473&partnerID=40&md5=64dfbfccd1549ae3567ac179975c378a}{On
  the hidden beauty of the proper orthogonal decomposition} 2~(5-6) (1991)
  339--352.
\newblock \href {http://dx.doi.org/10.1007/BF00271473}
  {\path{doi:10.1007/BF00271473}}.
\newline\urlprefix\url{https://www.scopus.com/inward/record.uri?eid=2-s2.0-0008458408&doi=10.1007%2fBF00271473&partnerID=40&md5=64dfbfccd1549ae3567ac179975c378a}

\bibitem{ABGA15}
L.~Alvergue, H.~Babaee, G.~Gu, S.~Acharya,
  \href{http://dx.doi.org/10.2514/1.J053295}{Feedback stabilization of a
  reduced-order model of a jet in crossflow}, AIAA Journal 53~(9) (2015)
  2472--2481.
\newblock \href {http://dx.doi.org/10.2514/1.J053295}
  {\path{doi:10.2514/1.J053295}}.
\newline\urlprefix\url{http://dx.doi.org/10.2514/1.J053295}

\bibitem{S10}
P.~Schmid, Dynamic mode decomposition of numerical and experimental data,
  Journal of Fluid Mechanics 656 (2010) 5--28.

\bibitem{KBBJ16}
J.~Kutz, S.~Brunton, B.~Brunton, J.~Proctor,
  \href{https://doi.org/10.1137/1.9781611974508}{Dynamic Mode Decomposition},
  Society for Industrial and Applied Mathematics, 2016.
\newblock \href {http://dx.doi.org/doi:10.1137/1.9781611974508}
  {\path{doi:doi:10.1137/1.9781611974508}}.
\newline\urlprefix\url{https://doi.org/10.1137/1.9781611974508}

\bibitem{LKB18}
K.~Laksari, M.~Kurt, H.~Babaee, S.~Kleiven, D.~Camarillo,
  \href{https://link.aps.org/doi/10.1103/PhysRevLett.120.138101}{Mechanistic
  insights into human brain impact dynamics through modal analysis}, Physical
  Review Letters 120~(13) (2018) 138101--.
\newblock \href {http://dx.doi.org/10.1103/PhysRevLett.120.138101}
  {\path{doi:10.1103/PhysRevLett.120.138101}}.
\newline\urlprefix\url{https://link.aps.org/doi/10.1103/PhysRevLett.120.138101}

\bibitem{VEKK19}
M.~Velegar, N.~B. Erichson, C.~A. Keller, J.~N. Kutz,
  \href{https://www.geosci-model-dev.net/12/1525/2019/}{Scalable diagnostics
  for global atmospheric chemistry using ristretto library (version 1.0)},
  Geoscientific Model Development 12~(4) (2019) 1525--1539.
\newblock \href {http://dx.doi.org/10.5194/gmd-12-1525-2019}
  {\path{doi:10.5194/gmd-12-1525-2019}}.
\newline\urlprefix\url{https://www.geosci-model-dev.net/12/1525/2019/}

\bibitem{Babaee:2017aa}
H.~Babaee, M.~Choi, T.~P. Sapsis, G.~E. Karniadakis,
  \href{http://www.sciencedirect.com/science/article/pii/S0021999117303364}{A
  robust bi-orthogonal/dynamically-orthogonal method using the covariance
  pseudo-inverse with application to stochastic flow problems}, Journal of
  Computational Physics 344 (2017) 303--319.
\newblock \href {http://dx.doi.org/https://doi.org/10.1016/j.jcp.2017.04.057}
  {\path{doi:https://doi.org/10.1016/j.jcp.2017.04.057}}.
\newline\urlprefix\url{http://www.sciencedirect.com/science/article/pii/S0021999117303364}

\bibitem{anand2005model}
M.~Anand, K.~Rajagopal, K.~Rajagopal, A model for the formation and lysis of
  blood clots, Pathophysiology of haemostasis and thrombosis 34~(2-3) (2005)
  109--120.

\bibitem{2015transport}
Z.~Li, A.~Yazdani, A.~Tartakovsky, G.~E. Karniadakis, Transport dissipative
  particle dynamics model for mesoscopic advection-diffusion-reaction problems,
  The Journal of chemical physics 143~(1) (2015) 014101.

\bibitem{KS05}
G.~E. Karniadakis, S.~J. Sherwin, Spectral/hp element methods for computational
  fluid dynamics, Oxford University Press, USA, 2005.

\bibitem{Babaee:2013ab}
H.~Babaee, S.~Acharya, X.~Wan,
  \href{http://dx.doi.org/10.1115/1.4025732}{Optimization of forcing parameters
  of film cooling effectiveness}, Journal of Turbomachinery 136~(6) (2013)
  061016--061016.
\newline\urlprefix\url{http://dx.doi.org/10.1115/1.4025732}

\bibitem{Babaee:2013aa}
H.~Babaee, X.~Wan, S.~Acharya,
  \href{http://dx.doi.org/10.1115/1.4025562}{Effect of uncertainty in blowing
  ratio on film cooling effectiveness}, Journal of Heat Transfer 136~(3) (2013)
  031701--031701.
\newline\urlprefix\url{http://dx.doi.org/10.1115/1.4025562}

\bibitem{Keromnes13}
A.~K{\'e}romn{\`e}s, W.~Metcalfe, K.~Heufer, N.~Donohoe, A.~Das, C.~Sung,
  J.~Herzler, C.~Naumann, P.~Griebel, O.~Mathieu, et~al., {A}n {E}xperimental
  and {D}etailed {C}hemical {K}inetic {M}odeling {S}tudy of {H}ydrogen and
  {S}yngas {M}ixture {O}xidation at {E}levated {P}ressures, Combust. Flame
  160~(6) (2013) 995--1011.

\bibitem{SR97}
L.~Shampine, M.~Reichelt, {T}he {MATLAB} {ODE} {S}uite, {SIAM} J. Sci. Comput.
  18~(1) (1997) 1--22.

\end{thebibliography}

\end{document}